\documentclass[journal]{vgtc}                
\ifpdf
  \pdfoutput=1\relax                   
  \usepackage{graphicx}                
  \DeclareGraphicsExtensions{.pdf,.png,.jpg,.jpeg} 
\else
  \usepackage{graphicx}                
  \DeclareGraphicsExtensions{.eps}     
\fi%

\graphicspath{{figures/}{pictures/}{images/}{./}} 

\usepackage{microtype}                 
\PassOptionsToPackage{warn}{textcomp}  
\usepackage{textcomp}                  
\usepackage{mathptmx}                  
\usepackage{times}                     
\usepackage{cite}                      
\usepackage{tabu}                      
\usepackage{booktabs}                  

\vgtccategory{Research}
\vgtcpapertype{algorithm/technique}

\usepackage{kotex}
\usepackage{subfigure}
\usepackage{caption}
\usepackage{algorithm, algpseudocode}
\usepackage{color}

\algblock{Input}{EndInput}
\algnotext{EndInput}
\algblock{Output}{EndOutput}
\algnotext{EndOutput}
\newcommand{\Desc}[2]{\State \makebox[10em][l]{#1}#2}

\title{Redirected Walking in Infinite Virtual Indoor Environment\\Using Change-blindness}

\author{June-Young Hwang, Soon-Uk Kwon, Yong-Hun Cho, Sang-Bin Jeon, and In-Kwon Lee}
\authorfooter{
\item
J.Y.Hwang, Y.H,Cho, S.U.Kwon, and S.B.Jeon are with the Department of Computer Science, Yonsei University, Seoul, Republic of Korea. E-mail: \{tlavotl1,maxburst88,kwonars,ludens0508\}@yonsei.ac.kr.
\item 
Y.H,Cho is with the Department of Computer Science, Yonsei University, Seoul, Republic of Korea. E-mail: maxburst88@gmail.com.
\item
I.K.Lee is with the Department of Computer Science, Yonsei University, Seoul, Republic of Korea. E-mail: iklee@yonsei.ac.kr.
}

\shortauthortitle{Hwang \MakeLowercase{\textit{et al.}}: Redirected Walking in Infinite Virtual Indoor Environment Using Change-blindness}

\abstract{
We present a change-blindness based redirected walking algorithm that allows a user to explore on foot a virtual indoor environment consisting of an infinite number of rooms while at the same time ensuring collision-free walking for the user in real space. This method uses change blindness to scale and translate the room without the user's awareness by moving the wall while the user is not looking. Consequently, the virtual room containing the current user always exists in the valid real space. We measured the detection threshold for whether the user recognizes the movement of the wall outside the field of view. Then, we used the measured detection threshold to determine the amount of changing the dimension of the room by moving that wall. We conducted a live-user experiment to navigate the same virtual environment using the proposed method and other existing methods. As a result, users reported higher usability, presence, and immersion when using the proposed method while showing reduced motion sickness compared to other methods. Hence, our approach can be used to implement applications to allow users to explore an infinitely large virtual indoor environment such as virtual museum and virtual model house while simultaneously walking in a small real space, giving users a more realistic experience. 
} 

\keywords{Virtual reality, redirected walking, space manipulation, change blindness}

\CCScatlist{ 
 \CCScat{K.6.1}{Management of Computing and Information Systems}%
{Project and People Management}{Life Cycle};
 \CCScat{K.7.m}{The Computing Profession}{Miscellaneous}{Ethics}
}

\teaser{
  \centering
  \subfigure[]{\includegraphics[width=0.45\linewidth]{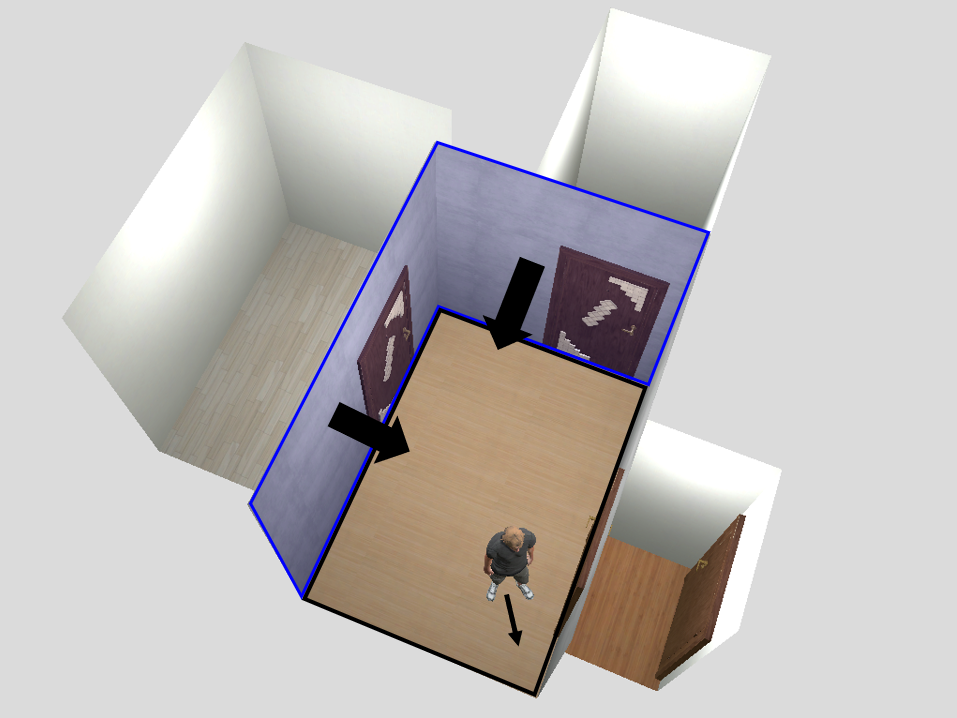}}
  \subfigure[]{\includegraphics[width=0.45\linewidth]{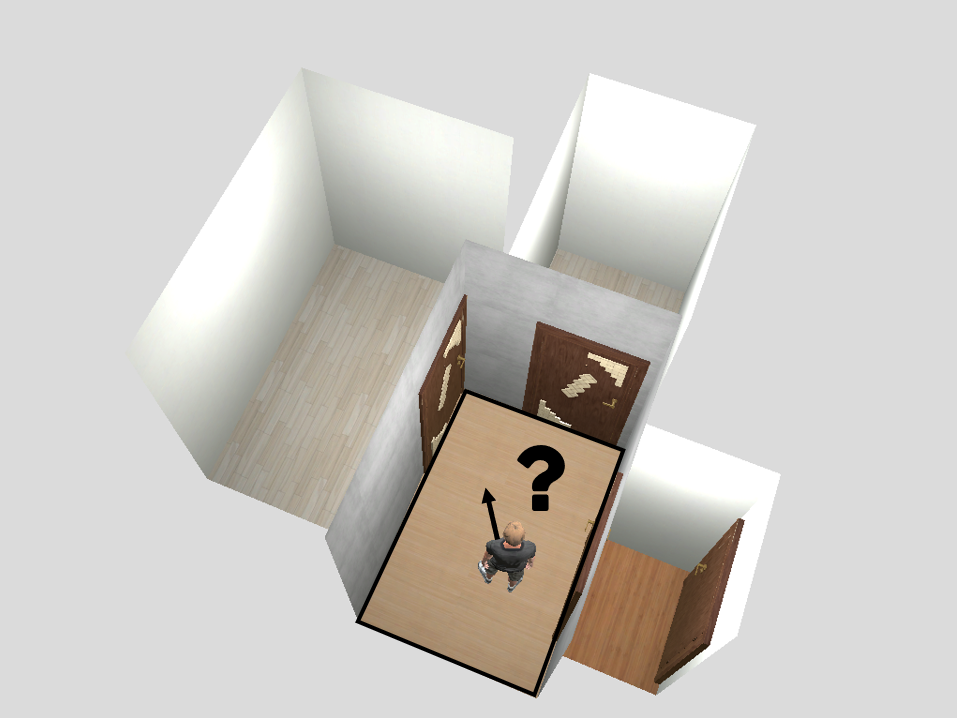}}
  \caption{Moving the walls of a room using change blindness: (a) before the wall moves, and (b) after the wall moves.}
    \label{fig:teaser}
}

\renewcommand{\manuscriptnotetxt}{}

\vgtcinsertpkg

\begin{document}


\firstsection{Introduction}

\maketitle

    Because locomotion in virtual reality affects a user's experience, various researches have studied locomotion techniques to provide a realistic experience\cite{bozgeyikli2016point,usoh1999walking, langbehn2015evaluation}. Among the locomotion techniques, actual walking induces higher immersion and natural and effective exploration for users\cite{suma2009evaluation, ruddle2009benefits, zanbaka2005comparison}. ReDirected Walking (RDW) has been studied as a technique that can overcome the difference between virtual and real space to explore a virtual environment through real walking. In particular, much research has been done on finely manipulating the user's scene in a range that the user is unaware of \cite{hodgson2013comparing, nescher2014planning, zmuda2013optimizing, thomas2019general}. On the other hand, some RDW studies\cite{suma2011leveraging, suma2012impossible, vasylevska2013flexible} have combined the change blindness phenomenon in which the user does not perceive the manipulation of the virtual space outside the field of view.

    However, the virtual environments applied in previous works of change blindness-based spatial manipulation in RDW are limited. In addition, it is necessary to measure the detection threshold to what extent humans are not aware of environmental changes outside of their field of view in a general indoor environment. Previous studies\cite{suma2012impossible,vasylevska2013flexible} have proposed a spatial manipulation algorithm using change blindness for a limited situation where the virtual space was divided into a manipulation area and a transition area, and measured the detection threshold to what extent the user was unaware of the spatial manipulation in that structure. Nevertheless, these methods cannot be applied to 
    some indoor environments such as a consecutively connected rooms without any corridors. Also, since the previous work procedurally generated virtual layout, these layout cannot be preserved. Therefore, users are more likely to be confused when trying to recall the structure of the virtual space they have experienced. Moreover, it is difficult to apply the measured detection threshold to other indoor environments because the detection threshold was measured in an impossible virtual environment consisting of two adjacent rooms that spatially overlap. 
    
    In this paper, we propose a novel RDW technique combined with change blindness-based spatial manipulation in 
    some indoor environments. Through this, change blindness-based redirection can be applied in more various indoor scenarios and furthermore, by combining the proposed method and the existing one, this method can be expanded to be used to a general virtual indoor environment. We generalize the virtual indoor environment as a set of rectangular rooms (which can be considered as rooms or corridors in indoor environment) and assume that doors could connect adjacent rooms. In this environment, our method moves the walls out of the user's field of view as Figure \ref{fig:teaser} and allows the user to walk the entire virtual space without colliding with the real space. Specifically, this technique applies `restore' and `compression' phases whenever the user enters a new room. The restore phase gradually moves the walls of the user's room until the room restores its original dimension and is centered in the user's physical space. The compression phase moves the walls of adjacent rooms into the real space so that all adjacent rooms the user is likely to enter next are completely contained within the real space.
    
    We defined out-of-view wall movement that can be applied in various virtual indoor environments and measured the detection threshold for its movement. Furthermore, we compared the results of user questionnaire by making the user perform the same virtual task in our method and other locomotion methods. As a result, the proposed method showed higher immersion, usability, and motion sickness reduction than Steer-To-Center (S2C) and showed a high sense of immersion and presence compared to Teleport, another conventional locomotion technique. 
    
    Based on our results, we claim the contribution of our research as follows. First, by combining spatial transformation using change blindness and RDW, we propose a new redirected walking method 
    applicable to diverse virtual indoor environments in which interconnected multiple rooms and corridors exist. Second, we defined the wall movement gain outside the user's field of view suitable in 
    various indoor environments and measured the detection threshold for the gain in empty room. Therefore, our method will provide a better VR experience to users by applying it to an infinite virtual indoor environment such as a virtual model house and museum consisting of multiple rooms and corridors 
    or games like room escape where the user stays in one room for a long time.

\section{Related work}
    \subsection{Redirected Walking (RDW)}
        RDW has been proposed to allow users to simultaneously walk in an infinitely wide and complex virtual space and a narrow, small, and limited real (physical) space. This method finely manipulates the user's view in a virtual environment to the extent that the user does not notice it. As a result, it induces the user's real movement and its virtual movement to be different from each other. For example, when the user moves forward in virtual space, this technique rotates the user's view to induce the user to circle in real space. A general overview of the RDW is provided in Suma et al.~\cite{suma2012taxonomy} and Nilson et al.~\cite{nilsson201815}. Hodgson et al.\cite{hodgson2013comparing} implemented the RDW concept, presented four steering algorithms, and conducted a performance comparison among the steering algorithms. Azmandian et al.~\cite{azmandian2015physical} compared the performance of these steering algorithms according to the actual space with various sizes and ratios. Also, they confirmed that it shows better performance if translation gain is added to the existing steering algorithm.
        
        Recently, several studies have shown better performance than the existing ones by combining a new concept with RDW. Sun et al.~\cite{sun2018towards} and Langbehn et al.~\cite{langbehn2018blink} have proposed an RDW that manipulates the user's scene between blinking and saccades. Thomas et al.~\cite{thomas2019general} proposed an RDW technique that guides users to a place far away from obstacles in real space by applying the concept of Artificial Potential Field (APF). Williams et al.~\cite{williams2021arc} proposed an Alignment-based Redirection Controller (ARC) that introduces alignment, representing the difference between the user's state in real space and virtual space, and steers the user toward decreasing alignment. Moreover, some studies \cite{lee2019real, strauss2020steering} have proposed techniques that combine RDW and reinforcement learning to apply optimal gains according to the states of user and environment.
    
    \subsection{Manipulating Virtual Space}
        Meanwhile, some studies have shown that RDW can be combined with the change blindness \cite{simons1997change} phenomenon, in which the user does not perceive the manipulation of the virtual space. Steinicke et al.~\cite{steinicke2010change} confirmed that change blindness occurred in a stereoscopic environment and showed the possibility of applying change blindness in VR environment. Suma et al.~\cite{suma2011leveraging} combined the change blindness phenomenon with RDW for the first time. In this study, they changed the locations of doors and corridors to avoid collisions between the user and the real space while the user was working on a computer in a virtual room. The work reported that most users left the room without knowing these changes and proceeded with the virtual environment. A follow-up study\cite{suma2012impossible} proposed a method to overlap the room with an adjacent room to extend its size while the user moves through the corridor. However, this technique cannot be applied unless a virtual environment is composed of a manipulation area and a translation area such as two adjacent rooms and a corridor. For example, it is impossible to explore the virtual environment with consecutive connecting rooms, one of the general indoor structures, without colliding with the real space.
        
        Vasylevska et al.~\cite{vasylevska2013flexible} first developed an algorithm that procedurally generates layout by computing room positions and connecting corridors' shapes. However, it is still applicable only in an environment where rooms and the corridor are separated, and it cannot fix the spatial layout, such as the location of rooms and doors and the shape of the corridor in advance. In addition, they did not measure the detection threshold of whether a user knew space manipulation whenever the layout of the virtual environment was procedurally generated and did not conduct the experiment on the quantitative measure of user questionnaire when applying this method.
    
    \subsection{Detection Thresholds}
         User's cognitive performance gets affected when a user perceives that the movement in virtual and real is different \cite{bruder2015cognitive, rietzler2018rethinking}. Hence, to make the user unaware of the distortion in the virtual environment, Steinicke et al.~\cite{steinicke2009estimation} defined how to distort the virtual space as gain. They presented a way to measure a detection threshold for the gain and its result. Based on this work, many studies have measured how the detection threshold of gain changes under various conditions. For example, Neth et al.~\cite{neth2012velocity} examined the curvature gain according to speed. Lucie et al.~\cite{kruse2018can} measured the change in translation gain with and without the user's feet, and Paludan et al.~\cite{paludan2016disguising} confirmed how visual density affects the rotation gain. Recently, some studies\cite{langbehn2017bending, langbehn2018blink, kim2021adjusting, cho2021walking} have defined new kinds of gains and measured the detection threshold for them.
        
        On the other hand, Suma et al.~\cite{suma2012impossible} measured the detection threshold for overlapping space for the first time in space manipulation. In detail, they measured the detection threshold for how much overlapping of rooms be distinguished by the user in a situation where two rooms are attached. In addition, they conducted a distance estimation task between two rooms for users according to the degree of overlap and confirmed the tendency for users to overestimate the distance between rooms even if they recognized the overlap. However, they measured the detection threshold only when two rooms are adjacent and expanded, so it is not easy to apply it to general structures different from the proposed structure. Vasylevska et al.\cite{vasylevska2015influence, vasylevska2017towards} also conducted an experiment to estimate the distance between two rooms according to the complexity of the corridor connecting the two rooms. Though, unlike Suma et al., they did not measure the detection threshold for overlapping virtual rooms and corridors, and it is also impossible to apply if the structure is not followed the proposed one. 
        
\begin{figure}[h]
    \centering
    \subfigure[]{\includegraphics[width=0.49\linewidth]{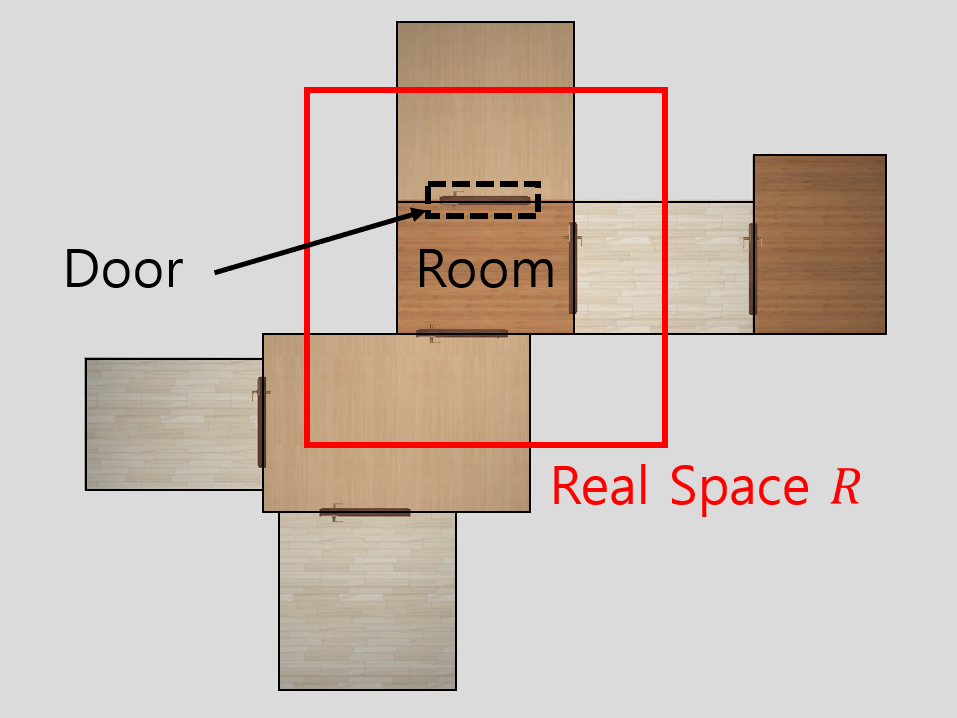}}
    \subfigure[]{\includegraphics[width=0.49\linewidth]{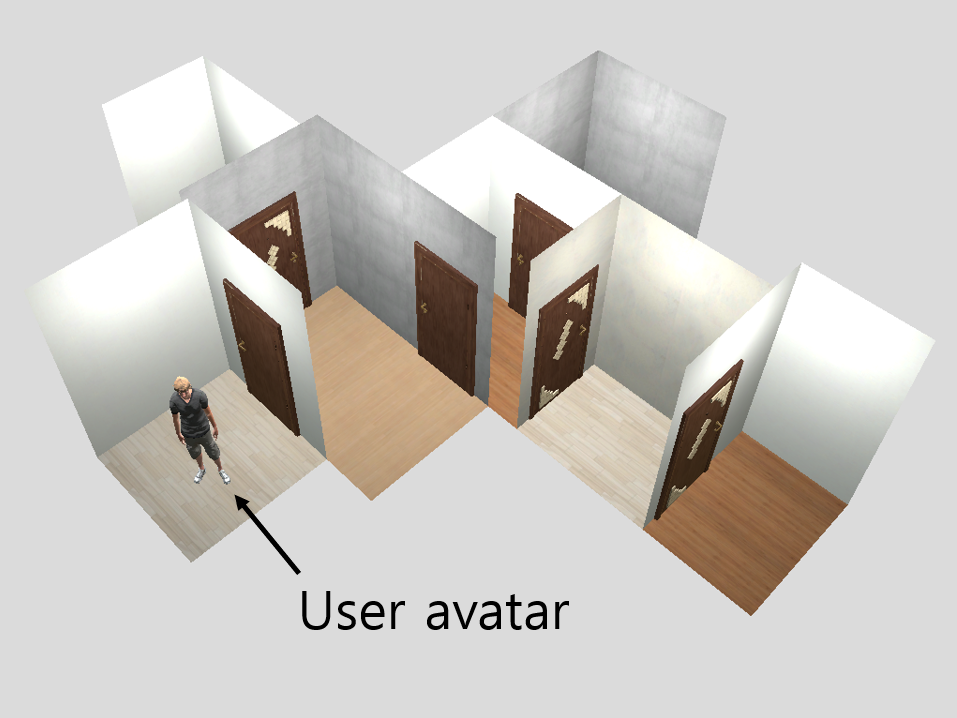}}
    \caption{An example of virtual and real environments: (a) the top view of the virtual environment consisting of the rooms connected with doors; the available real space is represented with a red rectangle, (b) the 3D perspective view of the virtual environment with virtual user avatar.} 
    \label{fig:env}
\end{figure}

\section{Method}
In this study, we propose a novel approach to space manipulation in RDW by moving the wall outside the user's field of view wherever the user is located in an arbitrary room in virtual indoor environments. Typically, an indoor environment is composed of rooms and corridors connecting them. Furthermore, if we consider the corridors are just another type of room, the general indoor environment can be represented as a set of the infinite number of adjacent rooms. Hence, we can assume the virtual indoor environment and the real space as Figure \ref{fig:env}. We assume the real space $R$ is a rectangle, and the virtual space $V$ consists of $n$ rooms, each of which is a rectangle with a size that can be contained entirely within the real space. Lastly, each room could be connected to another adjacent room by a door.

\begin{figure*}[t]
    \centering
    \subfigure[]{\includegraphics[width=0.3\textwidth]{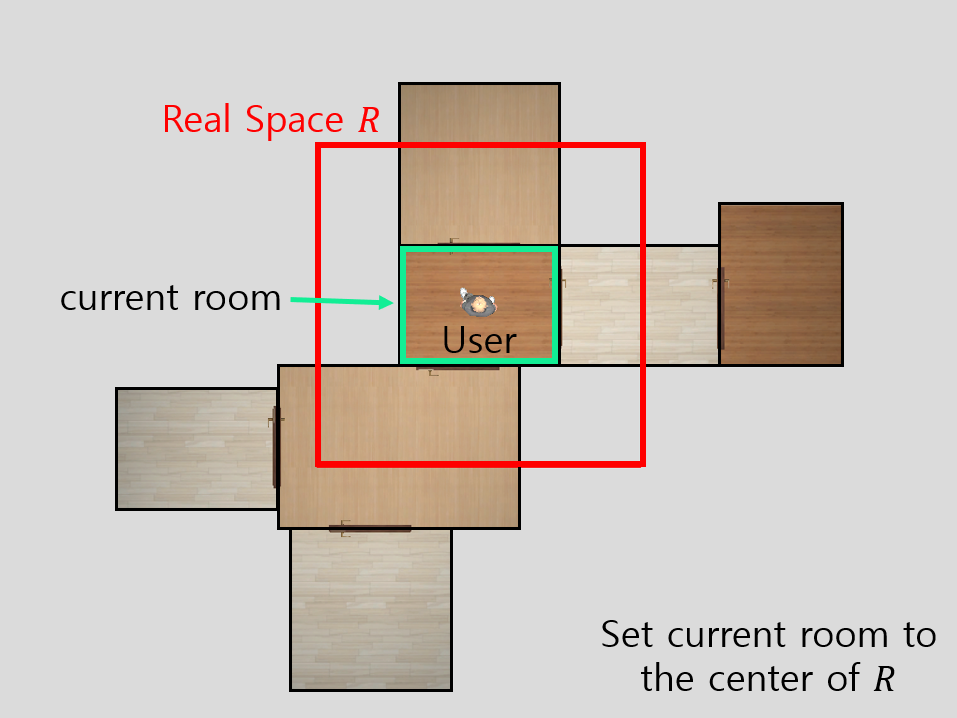}}
    \subfigure[]{\includegraphics[width=0.3\textwidth]{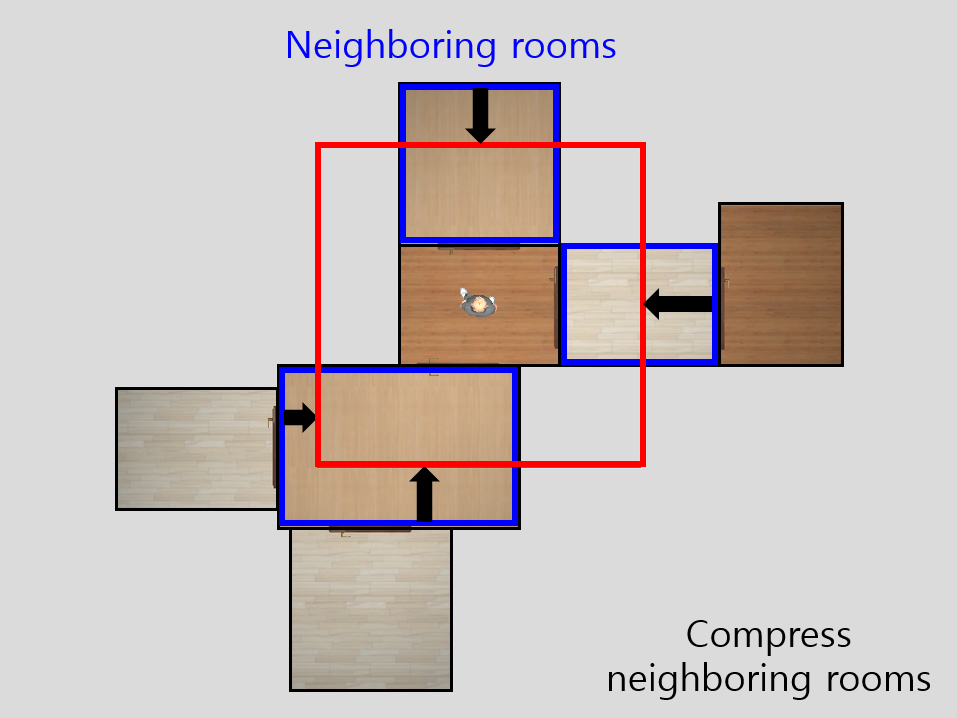}}
    \subfigure[]{\includegraphics[width=0.3\textwidth]{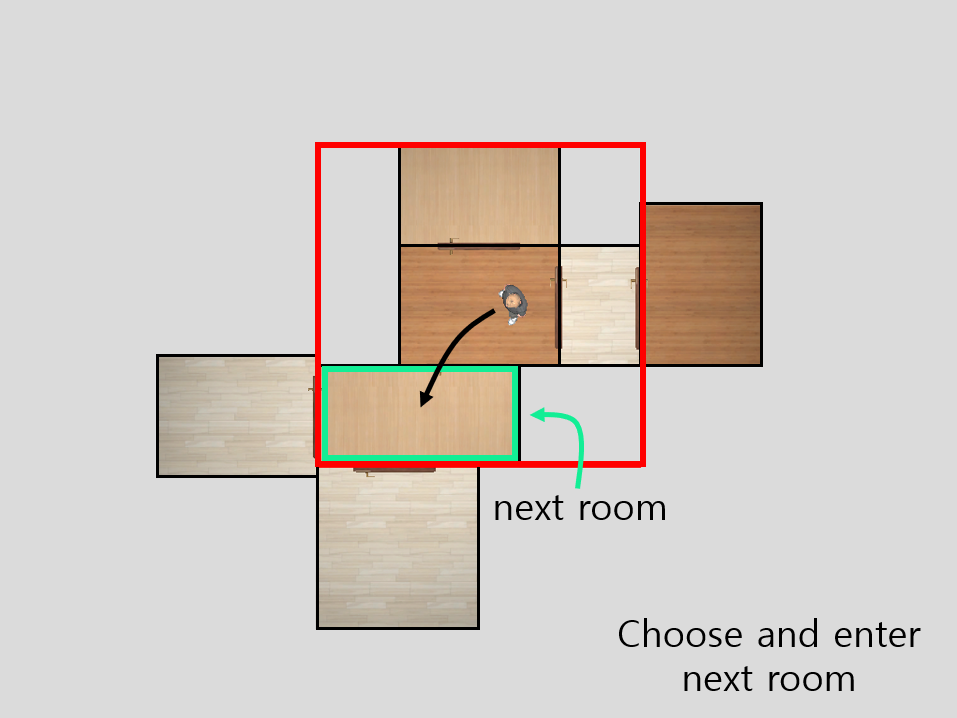}}
    \subfigure[]{\includegraphics[width=0.3\textwidth]{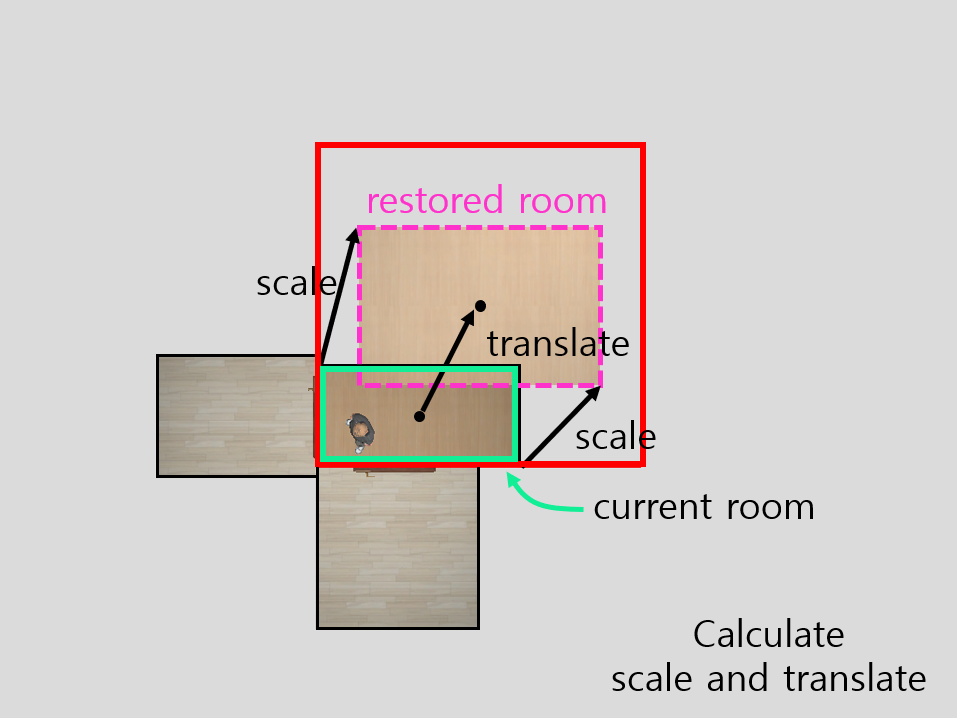}}
    \subfigure[]{\includegraphics[width=0.3\textwidth]{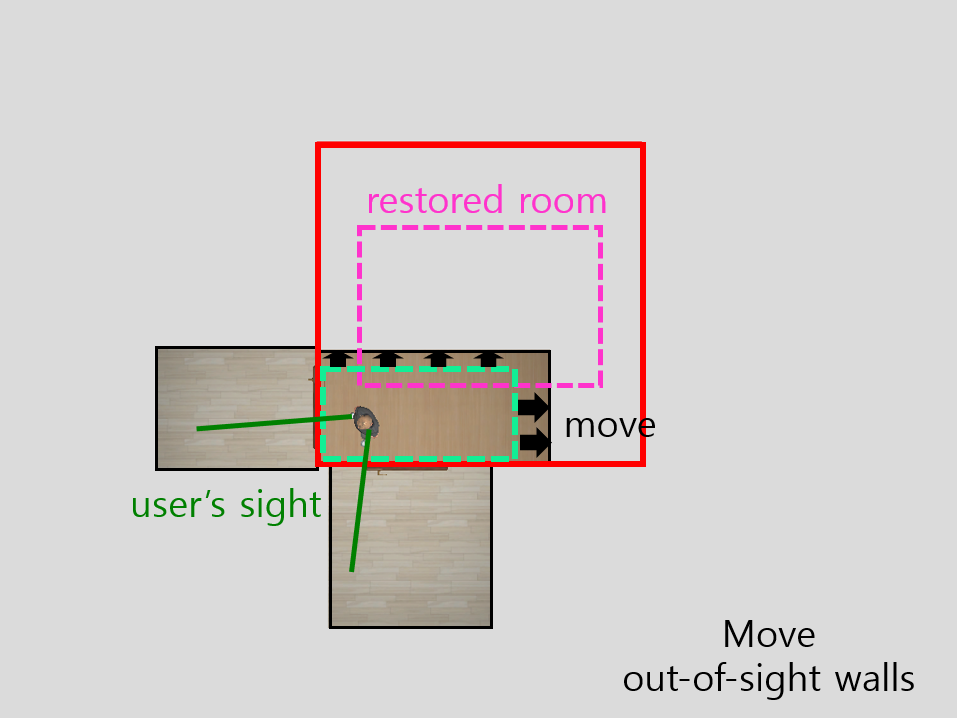}}
    \subfigure[]{\includegraphics[width=0.3\textwidth]{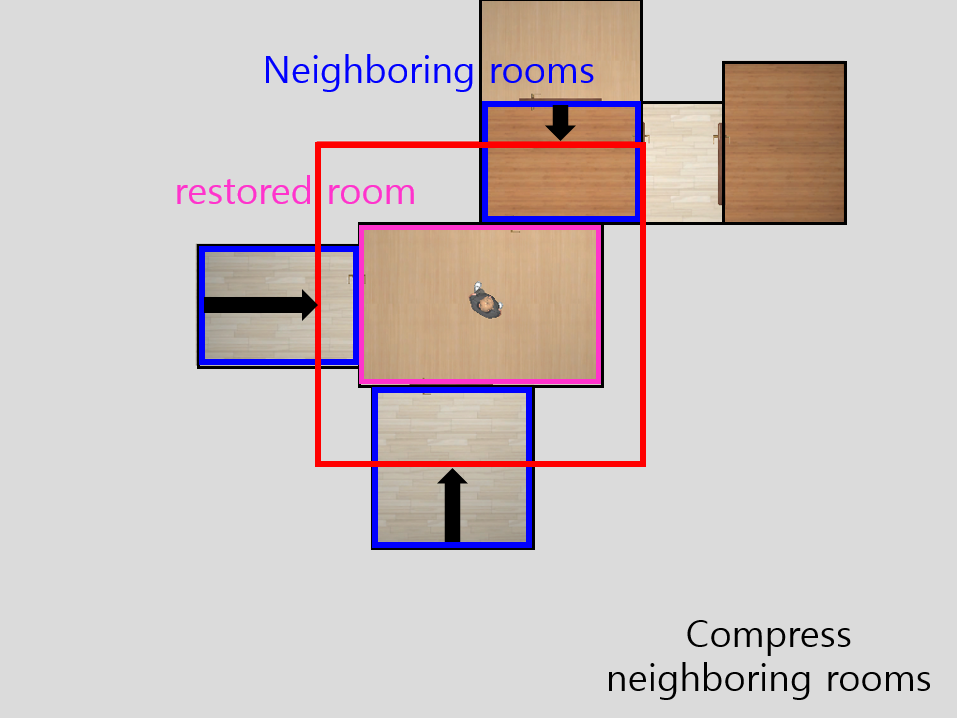}}
    \caption{The entire process of restore-compression.}
    \captionsetup{justification=centering}
    \label{fig:algo_step}
\end{figure*}

\begin{algorithm}[H]
\caption{Navigating Virtual Indoor Environment}\label{alg:nav}
    \begin{algorithmic}[1]
        \Input
            \Desc{Real space $R$:}{rectangle with center $O_c$}
            \Desc{Virtual space $V$:}{a set of rectangular rooms}
            \Desc{User $u$:}{an user}
        \EndInput%
        \Procedure{Navigate}{$R, V, u$}
            \State $r\gets$ a room that $u$ is initially located
            \State place $r$ at the center of $R$\label{alg:step1-1}
            \State \Call{Compress}{$R,r$}\label{alg:step1-2}
            \While{$u$ is navigating $V$}
                \If{$u$ enters a room $p \in V$}\label{alg:step2-1}
                    \State \Call{Restore-Compress}{$R,V,u,p$}\label{alg:step2-2}
                \EndIf
            \EndWhile
        \EndProcedure%
        \Statex
        \Procedure{Restore-Compress}{$R,V,u,p$}
            \State $D_o\gets$ original dimension of the room $p$
            \While{$u$ is staying in the room $p$}
                \State $O_p\gets$ current center of the room $p$
                \State $D_p\gets$ current dimension of the room $p$
                \State $s\gets$ scale from $D_p$ to $D_o$\label{alg:step3-1}
                \State $t\gets$ translation from $O_p$ to $O_c$\label{alg:step3-2}
                \If{$s\neq\mathbf{I}$ or $t\neq\mathbf{0}$}
                    \State \Call{Restore}{$u,p,s,t$}
                \Else
                    \State 
                    \Call{Compress}{$R,p$}
                   
                \EndIf
            \EndWhile
        \EndProcedure%
    \end{algorithmic}
\end{algorithm}

Under this assumption, we propose a general space manipulation algorithm for RDW as follows. As shown in Figure \ref{fig:algo_step}(a), \ref{fig:algo_step}(b), and line \ref{alg:step1-1}--\ref{alg:step1-2} in Algorithm \ref{alg:nav}, this method first set the room the user starts with to the center of the real space and then initializes the rooms by compressing all neighboring rooms of the initial room to the inside of the real space.

\begin{algorithm}[h]
\caption{Restoring Virtual Room}\label{alg:restore}
    \begin{algorithmic}[1]
        \Procedure{Restore}{$u,p,s,t$}
            \State $W\gets$ a set of walls in $p$ outside the field of view of $u$
            \For{$w\in W$}
                \State $w'\gets$ restored wall of $w$ computed with $s,t$\label{alg:step4}
                \State $d_w\gets$ vector from $w$ to $w'$
                \State $d_u\gets$ vector from $u$ to $w$
                \State $g_w\gets$ detection threshold of wall movement gain
                \If{$\left\Vert d_w \right\Vert\neq0$}
                    \State move $w$ towards $w'$ by $\left\Vert d_u \right\Vert \cdot g_w$\label{alg:step5}
                \EndIf{}
            \EndFor
        \EndProcedure
    \end{algorithmic}
\end{algorithm}

\begin{algorithm}[H]
\caption{Compressing All Neighboring Rooms}\label{alg:compression}
\begin{algorithmic}[1]
    \Procedure{Compress}{$R,p$}
        \State $Q\gets$ a set of adjacent rooms to $p$
        \For{$q\in Q$}
            \State $W_q\gets$ a set of walls in $q$
            \For{$w\in W_q$}
                \If{$w$ is completely out of $R$}
                    \State $w_r\gets$ the side of $R$ parallel and nearest to $w$\label{alg:step6-1}
                    \State $d_r\gets$ vector from $w$ to $w_r$
                    \State move $w$ towards $w_r$ by $\left\Vert d_r \right\Vert$\label{alg:step6-2}
                \EndIf{}
            \EndFor 
        \EndFor
    \EndProcedure
\end{algorithmic}
\end{algorithm}

Afterward, this method proceeds with the restore-compression phases whenever the user visits a new room (Figure \ref{fig:algo_step}(c) and line \ref{alg:step2-1}--\ref{alg:step2-2} in Algorithm \ref{alg:nav}). As shown in Figure \ref{fig:algo_step}(d), and line \ref{alg:step3-1}--\ref{alg:step3-2} in Algorithm \ref{alg:nav}, it calculates the scale from the current dimension of the room to its original dimension and the translation from the current center position of the room to the center of real space. The restore phase is performed if the calculated scale is not $\mathbf{I}$ or translation factor is not $\mathbf{0}$. In the restore phase, it first calculates the position to which each wall should be moved by combining the scale and translation obtained above (line \ref{alg:step4} in Algorithm \ref{alg:restore}). Next, it repeatedly moves the walls outside the user's field of view by applying the detection threshold of wall movement gain, which we define as a new type of gain to move the wall, until each wall reaches the corresponding position (Figure \ref{fig:algo_step}(e) and line \ref{alg:step5} in Algorithm \ref{alg:restore}). After the restore phase is completed, the compression phase compresses all neighboring rooms to the current user's room into the real space. For each wall in the neighboring room, the algorithm first finds the side of $R$ parallel and nearest to the wall. Then, it moves the wall towards that side of $R$ (Figure \ref{fig:algo_step}(f) and line \ref{alg:step6-1}--\ref{alg:step6-2} in Algorithm \ref{alg:compression}). As a result, it is possible to walk the virtual space where the infinite number of rooms are interconnected without colliding with real space by repeating these restoration-compression phases.

This method always abruptly compresses the neighbor's room. Also, since only the room the user located is restored to its original size while the user moves and looks around the current room, most room does not have the original size but is compressed. Hence, the user can feel this extreme distortion every time he enters a room. However, when a user stays in one room for a long time, such as a model house, a museum, and a room escape game, it is rare that the user perceives the compression of the room during the entire experience time since the user doesn't have to move much between each room. That is, in such applications, it can be expected that the user can explore the virtual environment through actual walking while offsetting the above disadvantages to some extent by applying the proposed method.

\section{Experimental Result}

    \subsection{Experiment 1: Wall Movement Gain}
        Our method moves out-of-view walls to the extent that a user does not perceive the movement. Therefore, it is necessary to measure the detection threshold for the movement of the wall behind the user. Also, change blindness is a phenomenon in which humans do not realize these changes even though they see changes in visual stimuli. To the best of our knowledge, 
        there are no studies that analyze the detection threshold of change blindness in a general condition. Also, only one of change blindness-based RDW researches measured the detection threshold and only measured it in a specific case when two adjacent rooms overlap. Consequently, 
        we measure the detection threshold of the wall movement in a room to apply that to our proposed algorithm. to estimate the detection threshold for wall movement, we made two assumptions. First, the more significant the change in the user's scene, the less likely change blindness will occur, hence the more users notice the changes. Second, many factors affect the user's scene, such as the user's viewing angle, the user's current state, the texture of the wall, the size of the room, and the relative position of the furniture in the room. However, among them, we suppose the distance between the user and the moving wall may significantly impact the visual changes in our environment. Hence, we evaluated the detection threshold of wall movement as three types of distance: Short (1m), Middle (2m), and Long (3m).
        
        We defined a new type of gain about wall movement applied in a typical virtual room. Precisely, we assume $T_b$ as the shortest distance between the user and the wall before the wall moving, and $T_a$ as the shortest distance between the user and the wall after the wall moving. And then, we define a wall movement gain as $g_w=T_a / T_b$. That is, if $g_w<1$, the wall moves closer to the user. $g_w=1$ means the wall does not move. if $g_w>1$, it means the wall moves away from the user. To estimate the detection threshold for wall movement gain, we conducted a preliminary experiment 
        and modified the gains to appropriate values. As a result, we manipulated the wall movement gain $g_w$ at intervals of 0.1 from 0.8 to 1.2 for Long and Middle, and 0.05 intervals from 0.9 to 1.1 for Short when applying the gain to users. We applied each wall movement gain three times, collected the user responses, and used the average of the user's responses in three times as the result. To summarize, we performed measurements of 45 $(3 \ \mbox{distances} \times 5 \ \mbox{gains} \times 3 \ \mbox{repeats}) $ times for each user, and the sequence was random.

\begin{figure}[h]
            \centering
            \subfigure[]{\includegraphics[width=0.32\linewidth]{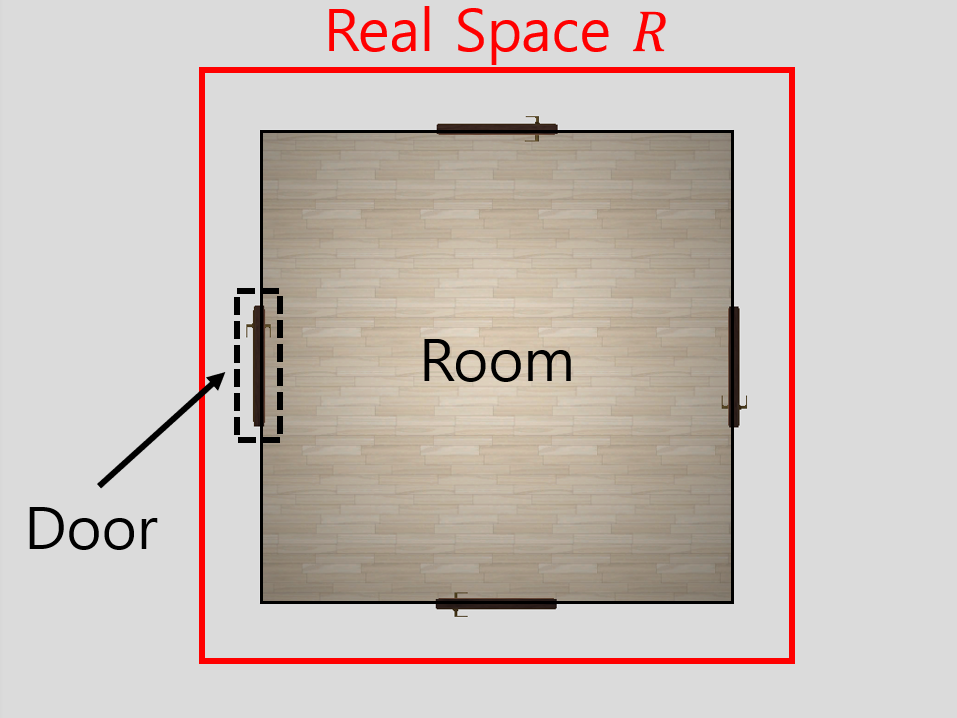}}
            \subfigure[]{\includegraphics[width=0.32\linewidth]{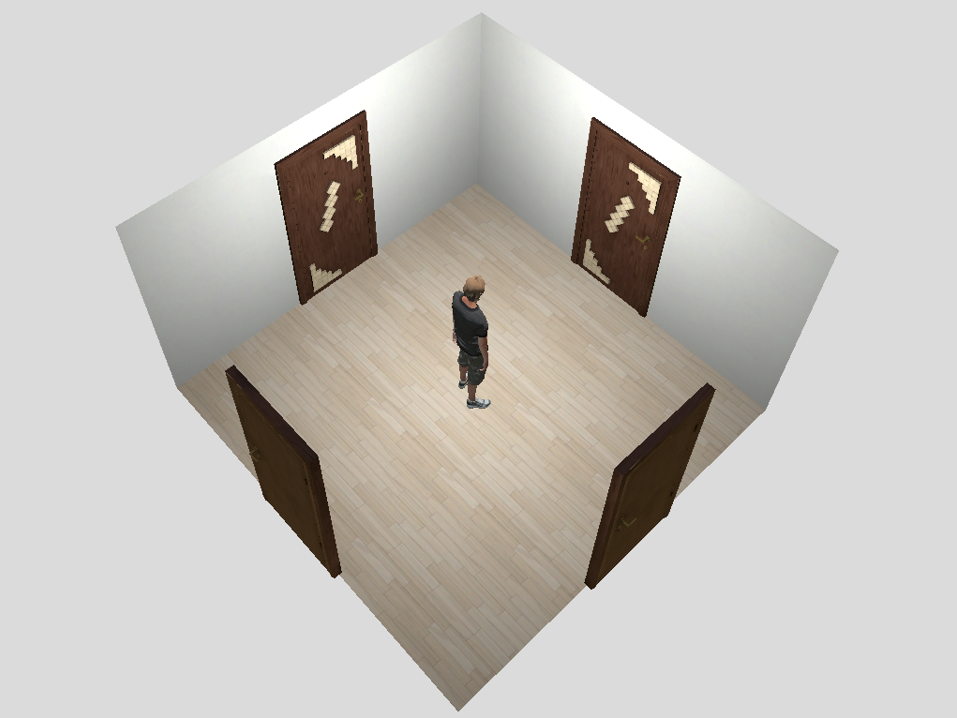}}
            \subfigure[]{\includegraphics[width=0.32\linewidth]{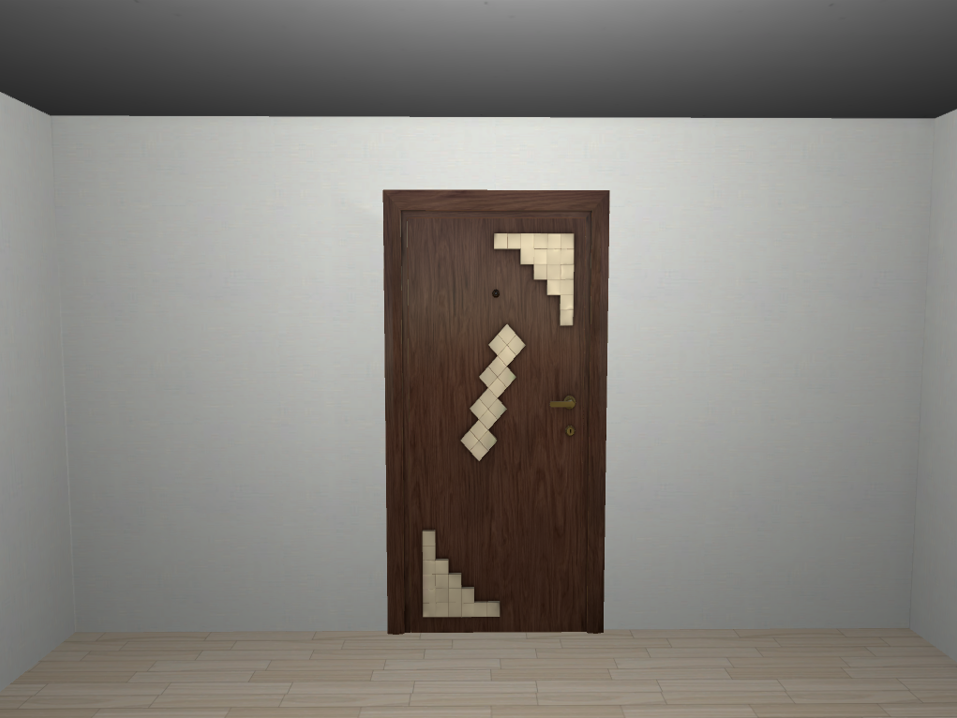}}
            \caption{The environment of Experiment 1: (a) the top view, (b) the 3D perspective view, (c) one of the user's views.}
            \label{fig:exp1_env}
        \end{figure}

        \paragraph{\bf Setup}
            We created a virtual space using Unity 3D as shown in Figure \ref{fig:exp1_env}(a), (b) and (c). The experimental space consists of a single rectangular room that fits perfectly within the real space of the room-scale size $4m \times 4m$. The virtual space has a light on the ceiling and doors located in the center of each wall. The texture of each wall was set to a monotone color. We used HTC Vive Pro HMD as the VR equipment for the experiment. The simulation program was created and run on a PC with an i7-8700 CPU, 16 GB of RAM, and an Nvidia Geforce 1070 GPU.
        
        \paragraph{\bf Participants}
            20 people (12 men and 8 women) participated in our experiment; they were between the ages of 20 and 30 years (mean = 23.4, SD= 3.1), and were students or office workers. 6 participants had no VR experience, 9 participants had VR experience once or twice, and 5 reported they had experienced VR more than three times. 14 people wore glasses and lenses, and 15 people were right-handed. After the experiment was over, all participants got incentives of 10\$.
            
\begin{figure}[t]
            \centering
            \subfigure[]{\includegraphics[width=0.49\linewidth]{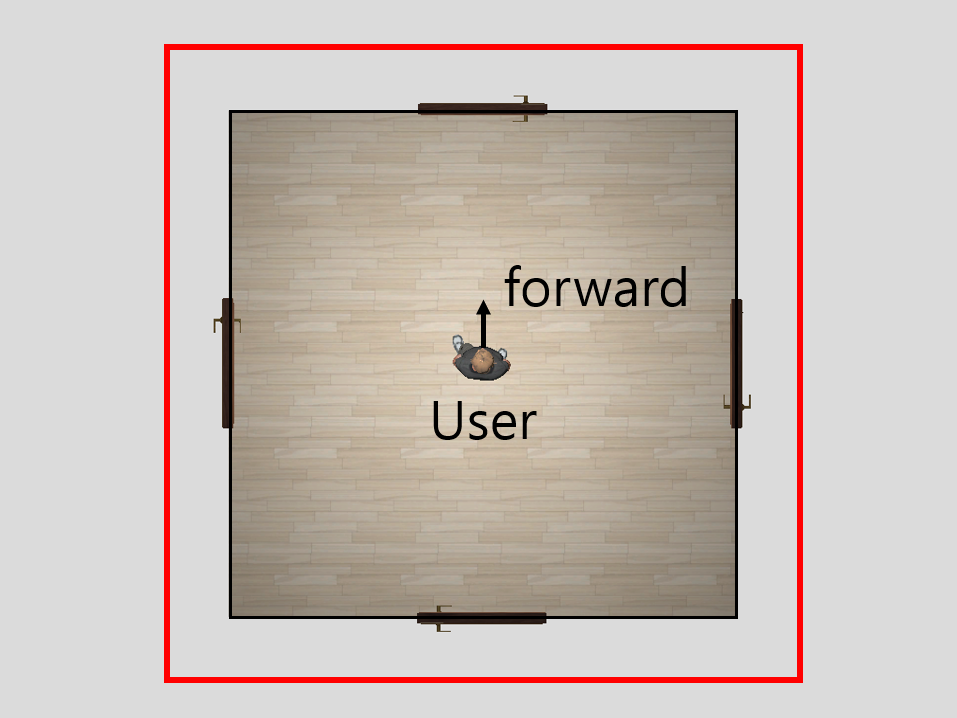}}
            \subfigure[]{\includegraphics[width=0.49\linewidth]{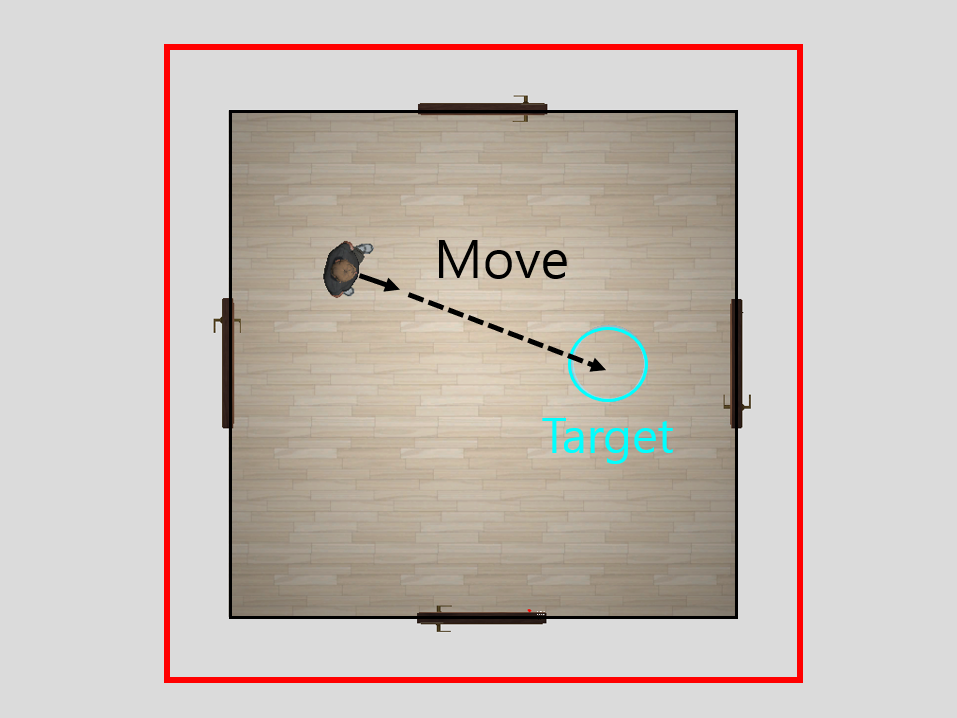}}
            \subfigure[]{\includegraphics[width=0.49\linewidth]{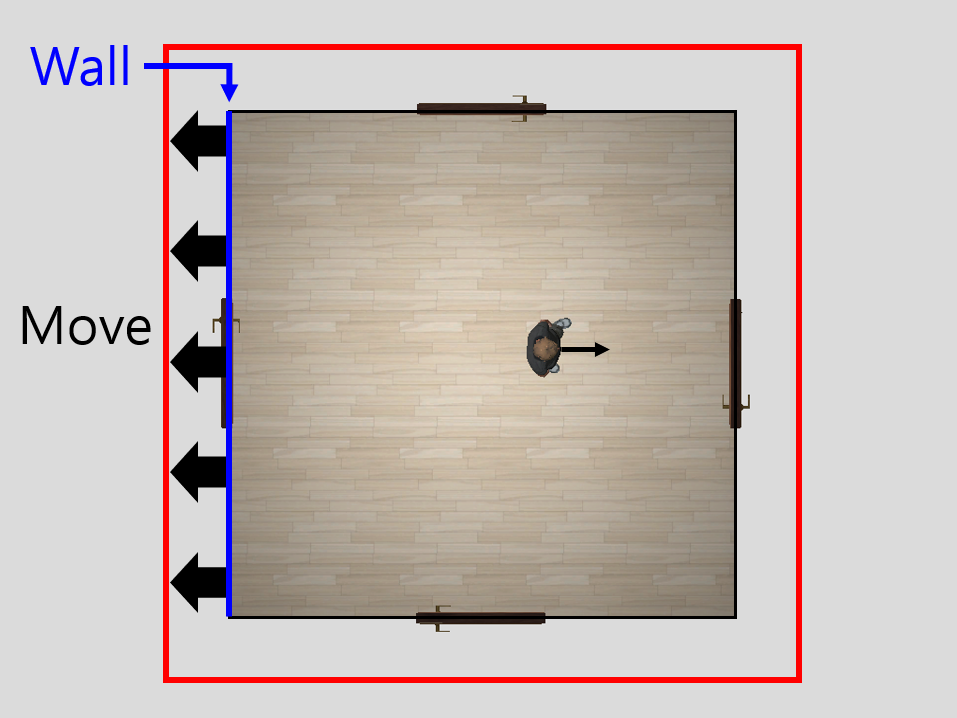}}
            \subfigure[]{\includegraphics[width=0.49\linewidth]{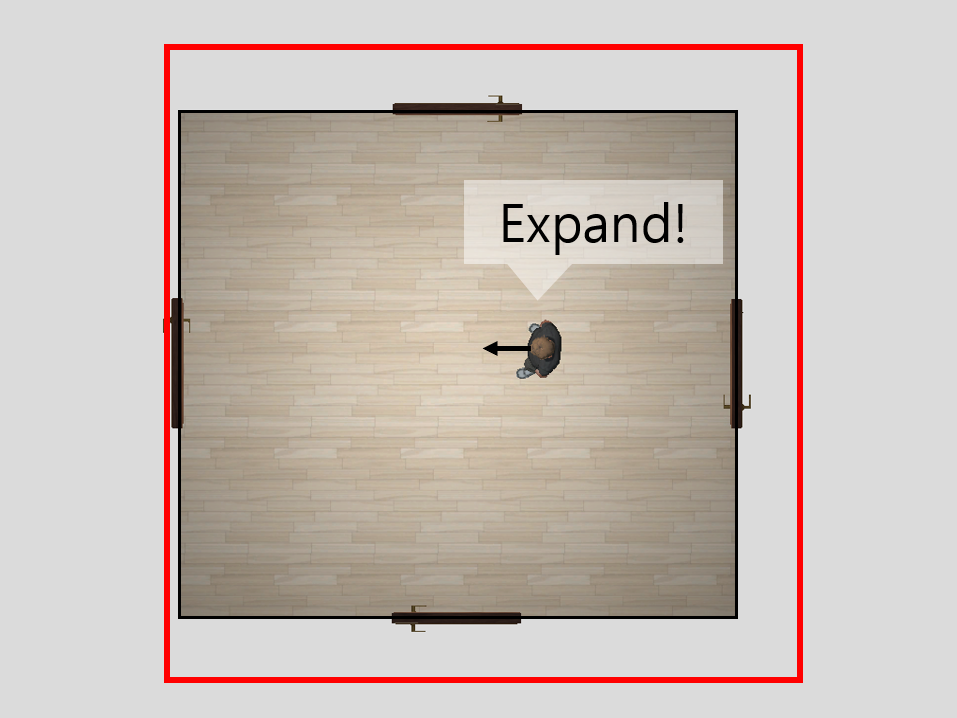}}
            \caption{The entire procedure of Experiment 1: (a) the user initially locates in the center of the room, (b) the user moves to the target, (c) the wall behind the user moves, (d) the user turns around and answers whether the room is expanded or shrinked}
            \label{fig:exp1}
        \end{figure}

        \paragraph{\bf Procedure}
            We informed all participants that they could take a break or stop if they felt motion sickness during the experiment. Afterward, we conducted a simple test session to prevent the learning effect by making the users adapt to VR and tasks in the experiment. After the test session was over, we moved participants to the center of the actual space and began the experiment. The user started at the center of the virtual room and look around their room for 10 seconds (Figure \ref{fig:exp1}(a)). Then, as shown in Figure \ref{fig:exp1}(b), the user was guided to move the specific position according to the distance to which the wall movement gain is applied. After the user reached the that position and was looking at the indicated door, we moved the wall behind the user by applying a wall movement gain (Figure \ref{fig:exp1}(c)). Then, the user turned around and observed the wall as shown in Figure \ref{fig:exp1}(d). Finally, the user asked, "Did the space get larger or smaller?" and they chose one of two answers, which are "It got larger" and "It got smaller", to the question. After answering, the user moved to the center while the screen blinks and move to the next case for other gains.
        
        \begin{figure*}[t]
            \centering
            \subfigure[]{\includegraphics[width=0.33\linewidth]{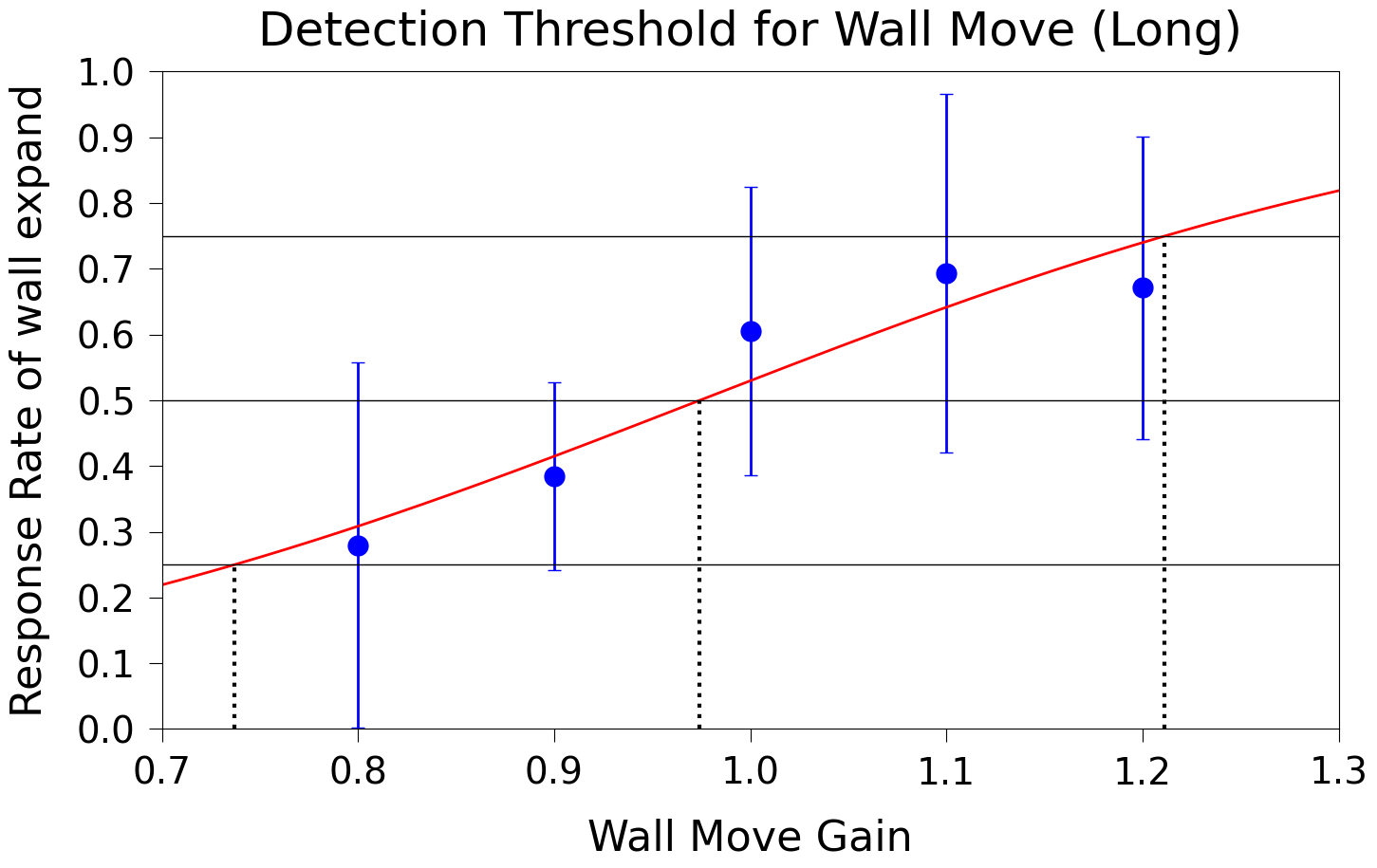}}
            \subfigure[]{\includegraphics[width=0.33\linewidth]{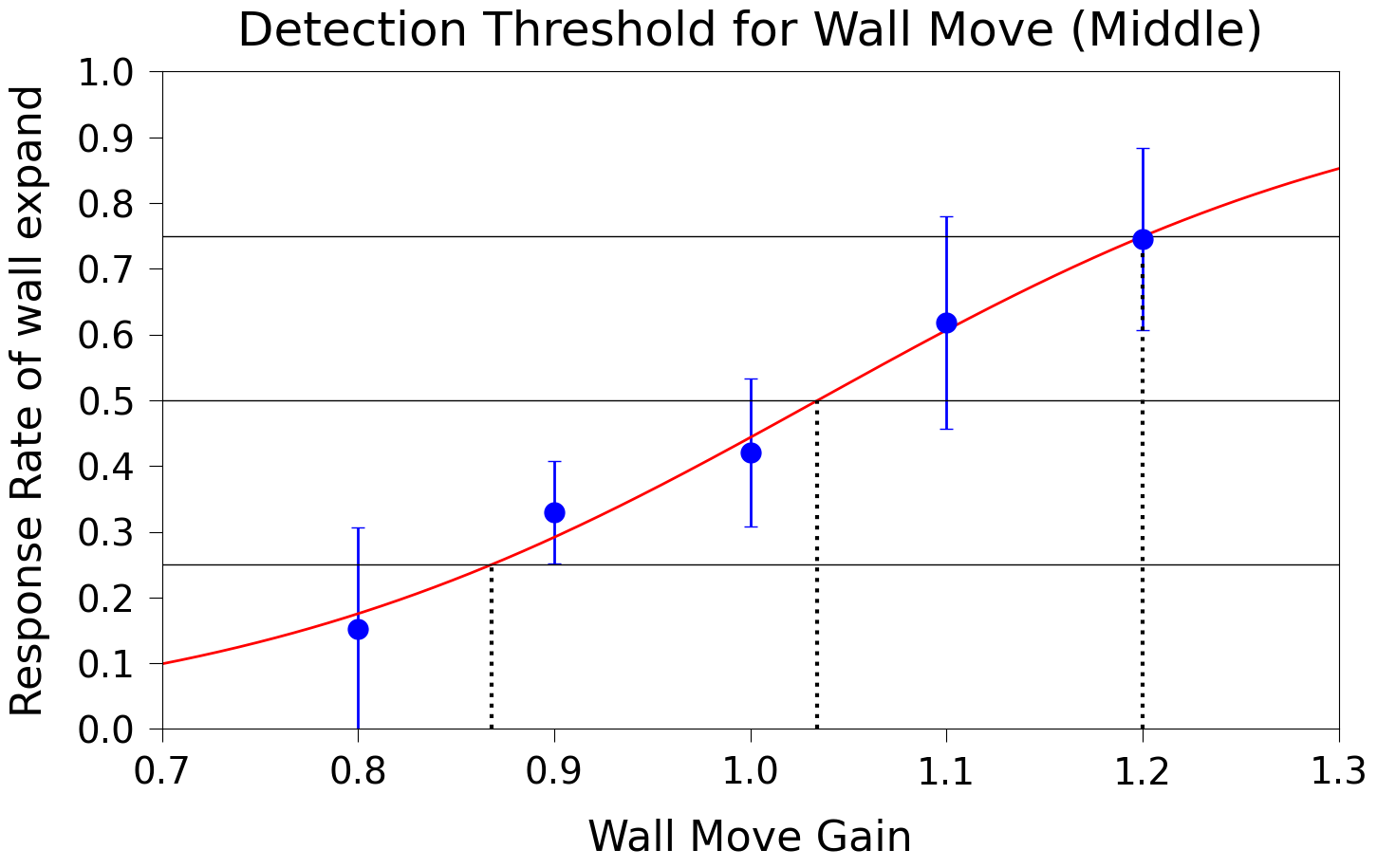}}
            \subfigure[]{\includegraphics[width=0.33\linewidth]{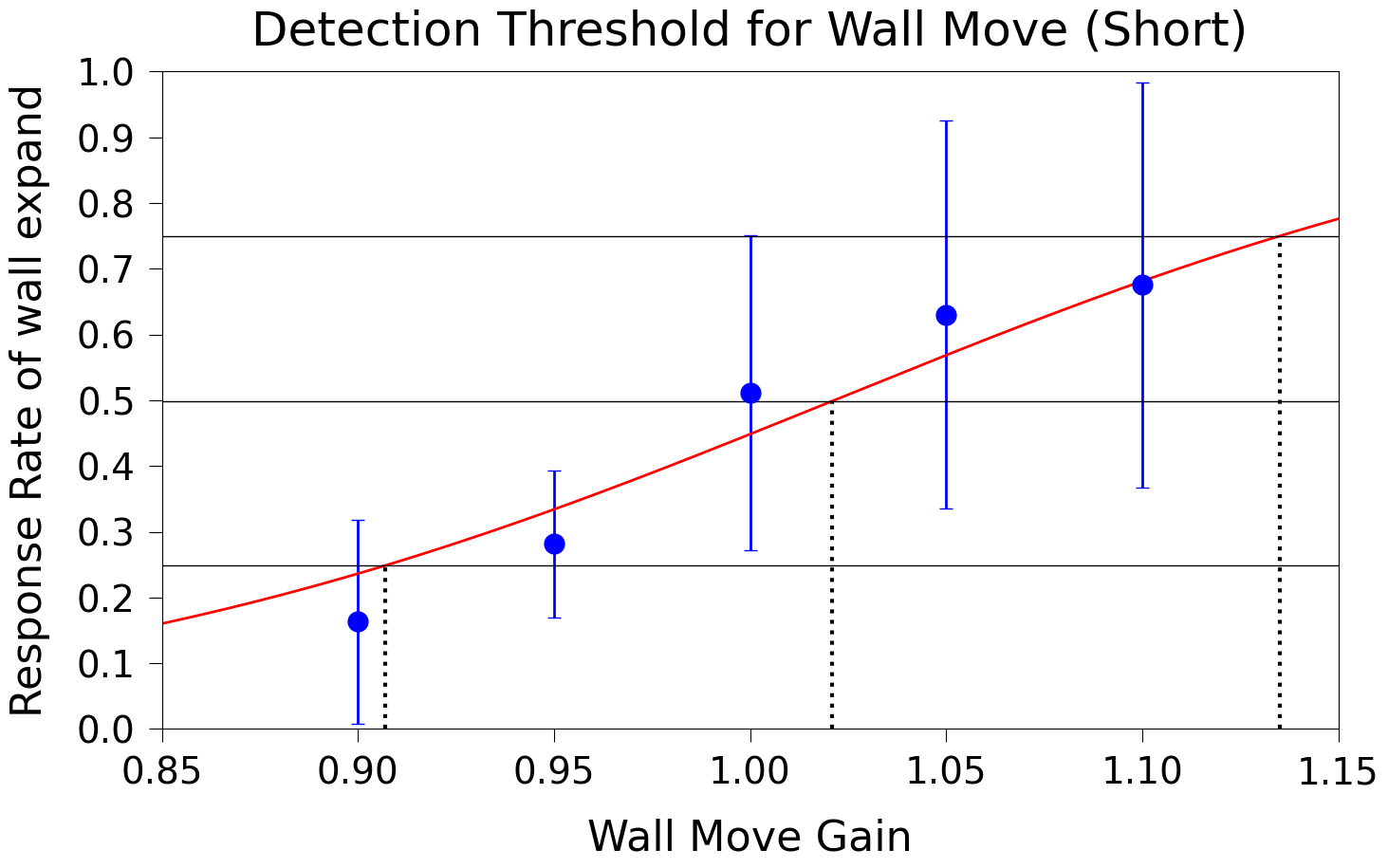}}
            \caption{Detection Threshold for Wall movement gain (DTW): (a) DTW in Long (3m away from the user), (b) DTW in Middle (2m away from the user), (c) DTW in Short (1m away from the user)}
            \label{fig:exp1_result}
        \end{figure*}

        \paragraph{\bf Detection Threshold}
            We used the two-alternative forced method during the experiment, such as "Did the space get larger or smaller?". Two-alternative forced choice is a question-answering method in which one cannot choose a neutral answer when two opposing answers are given but must be forcibly select one of them. We estimated the detection threshold by fitting the user's response obtained through this process to the psychometric function $f(x)=1/\left ( 1+e^{ax+b} \right )$. The point corresponding to 0.5 in the psychometric function is called Point of Subjective Equality (PSE), where the subject does not know whether the wall has moved at all. On the other hand, as the gain closer to 0 or 1 is applied, the user accurately recognizes whether the space is smaller or bigger because a significantly larger gain is applied. In Steinicke et al.~\cite{steinicke2009estimation}, they took the range of detection threshold from the lower point corresponding to 0.25, which is the midpoint between PSE and 0, to the upper point corresponding to 0.75, which is the midpoint between PSE and 1. Therefore, we fitted the response rate of expansion by the users for each wall movement gain to a psychometric function and took the range of wall movement gain corresponding to 0.25 and 0.75 in the estimated psychometric function as the range of detection threshold. 

        \paragraph{\bf Result}
            Table \ref{table:exp1_result} and Figure \ref{fig:exp1_result} summarize the result of the detection threshold for each distance between the user and the moving wall. In Figure \ref{fig:exp1_result}, $x$-axis represents the wall movement gain, and $y$-axis represents the response rate of expansion. The black dashed lines perpendicular to the $x$-axis represent the wall movement gain at the lower threshold (25\%), PSE (50\%), and upper threshold (75\%) points, respectively, from left to right. In the Long condition, the detection threshold ranged from 0.737 to 1.211, and the PSE was 0.974 (Figure \ref{fig:exp1_result}(a)). If we represent the detection threshold for the Long condition as meters, the user does not notice the changes until the wall moves toward the user by approximately 0.8m or moves away from the user by approximately 0.6m. In the Middle condition, the detection threshold ranged from 0.868 to 1.2, and the PSE was 1.03. Lastly, in the Short condition, the detection threshold ranged from 0.899 to 1.145, and the PSE was 1.02.

        \begin{table}[h]
            \centering
            \begin{tabular}{lccc}
            \toprule
            Distance from wall & Lower & PSE & Upper \\
            \midrule
            Long (3m) & 0.737 & 0.974 & 1.211\\
            Middle (2m) & 0.868 & 1.030 & 1.200\\
            Short (1m) & 0.899 & 1.020 & 1.145\\
            \bottomrule
            \end{tabular}
            \caption{Detection threshold for wall movement gain according to the distance between user and wall.}
            \label{table:exp1_result}
        \end{table}

    \subsection{Experiment 2: Navigating Indoor Environment}
        To check the effect of our method on user experience compared to other locomotion techniques, we conducted a live-user experiment to examine user questionnaires for our technique (Ours), S2C~\cite{razzaque2005redirected}, and Teleport~\cite{bozgeyikli2016point}. 
        The reason we did not compare it with previous spatial manipulation studies is that these studies~\cite{suma2011leveraging, suma2012impossible} only presented the technique and did not suggest an algorithm for how to apply it. In addition, since our study presents an algorithm that can be applied to indoor environments different from early study~\cite{vasylevska2013flexible}, it is impossible to compare performance with previous change blindness-based studies in the same environment. We formulated the following hypotheses {\bf H1-1},{\bf H1-2},{\bf H1-3},{\bf H1-4},{\bf H2-1},{\bf H2-2},{\bf H2-3},{\bf H2-4} and tested these hypotheses through statistical analysis. For each locomotion technique, we used SUS~\cite{brooke1996sus} to indicate usability, SUSPQ~\cite{usoh2000using} to indicate presence, IEQ~\cite{jennett2008measuring} to indicate immersion, and SSQ~\cite{kennedy1993simulator} to indicate the degree of motion sickness. We also collected qualitative feedback from users by asking some additional questions after the experiment. We excluded some questions of IEQ not suitable for our experiments, such as ``How much did you want to win the game?" and ``To what extent did you enjoy the graphics and imagery?". When applying the proposed method, we applied the detection threshold for wall movement gain obtained in Experiment 1. We designed an experiment with a within-subject design to offset baseline differences for each user. Also, the order of all conditions was counterbalanced.

        \begin{itemize}
            \item {\bf H1-1}, {\bf H1-2}, {\bf H1-3}, {\bf H1-4}: the score of SUS, SUSPQ, IEQ, and SSQ will be better in Ours than S2C, respectively. 
            \item {\bf H2-1}, {\bf H2-2}, {\bf H2-3}, {\bf H2-4}: the score of SUS, SUSPQ, IEQ, and SSQ will be better in Ours than Teleport, respectively. 
        \end{itemize}
        
\begin{figure}[h]
            \centering
            \subfigure[]{\includegraphics[width=0.32\linewidth]{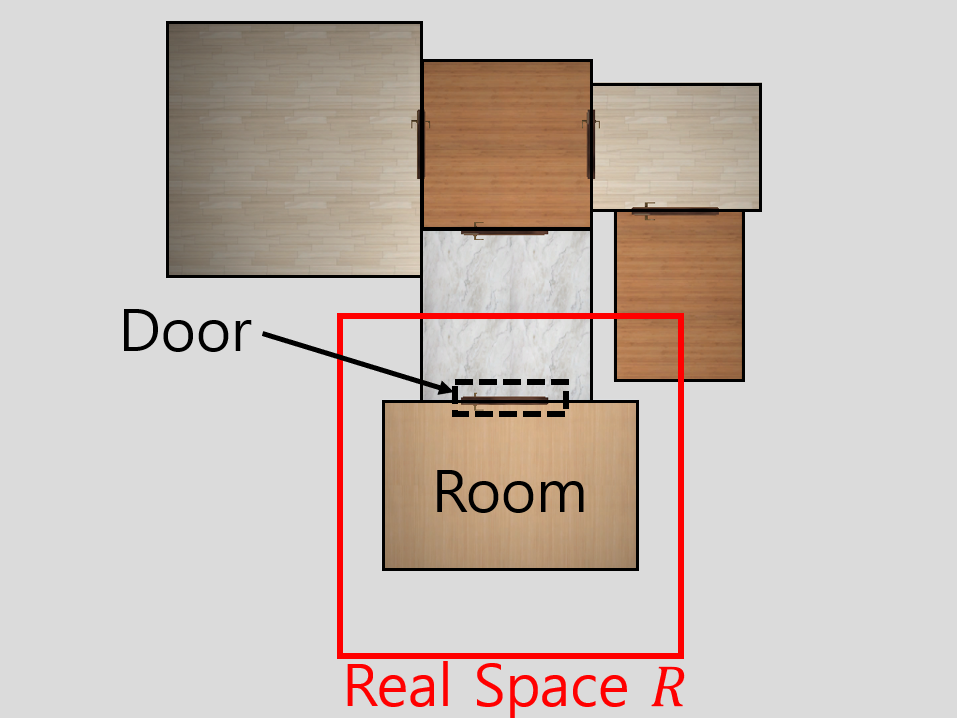}}
            \subfigure[]{\includegraphics[width=0.32\linewidth]{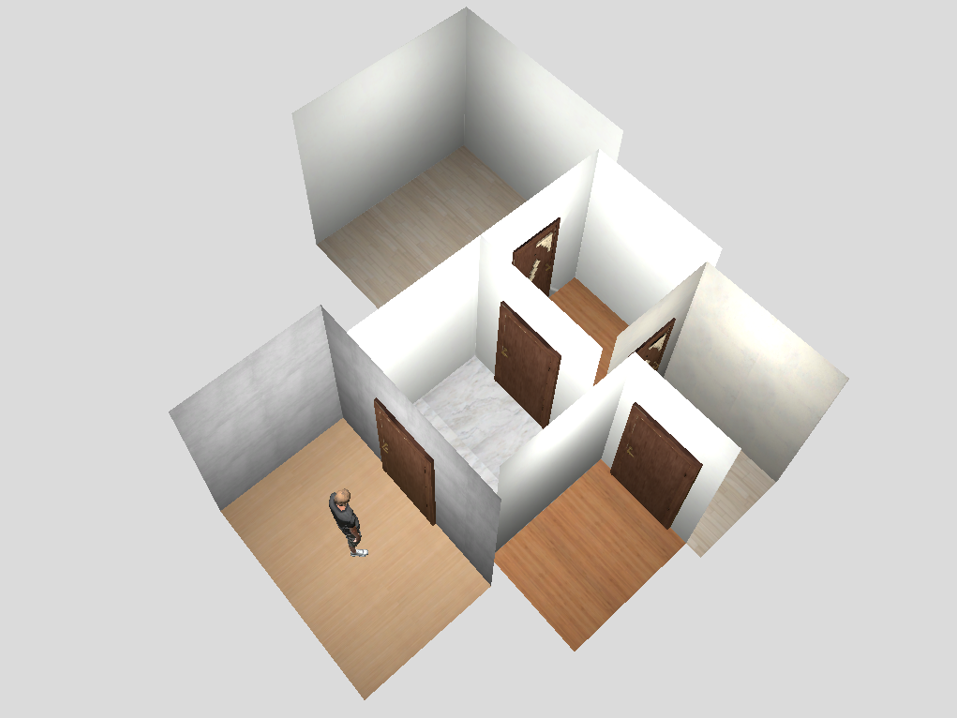}}
            \subfigure[]{\includegraphics[width=0.32\linewidth]{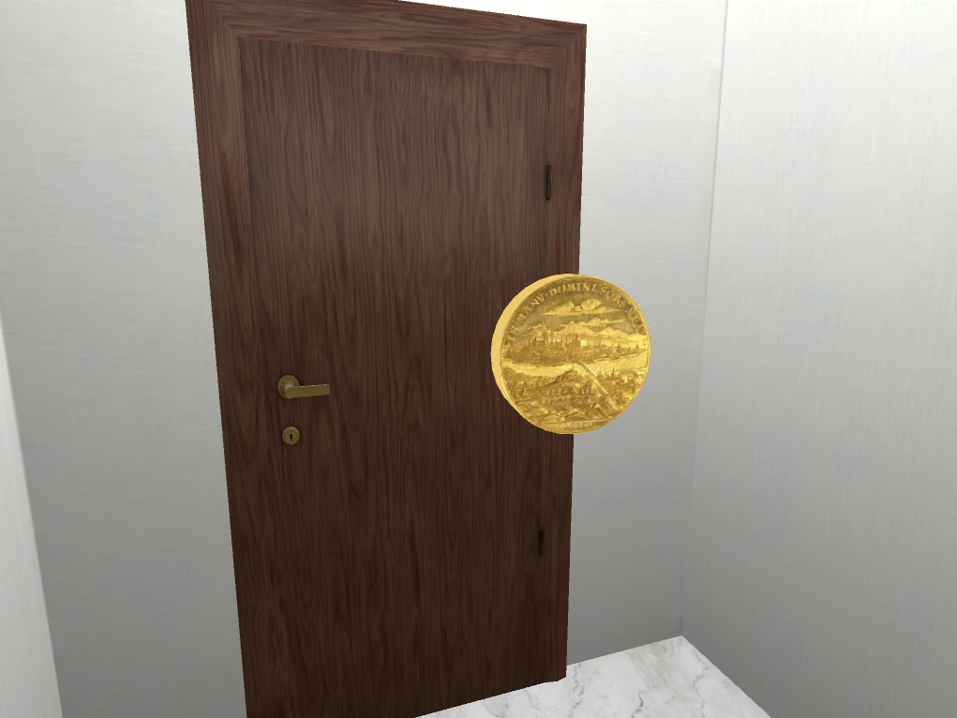}}
            \caption{The environment of Experiment 2: (a) the top view, (b) the 3D perspective view, (c) one of the user's views.}
            \label{fig:exp2_env}
        \end{figure}

        \paragraph{\bf Setup}
            We created a virtual space in Unity 3D as shown in Figure \ref{fig:exp2_env}(a)(b)(c). As the assumption mentioned in Section 3, the virtual space consists of several rectangular rooms, each of which can fit entirely into the real space. Some rooms are connected through a door, and the user can move to the connected room by pulling the door handle to open the door. Each room in the virtual space has a light on the ceiling and monotone color texture of the walls, but a different color is applied to each room to distinguish them. The VR equipment and computer used for the experiment are the same as in Experiment 1.
            
        \paragraph{\bf Participants}
            29 people (18 men and 11 women) participated in our experiment; they were between the ages of 20 and 30 years (mean = 25.1, SD= 3.4), and were students or office workers. 9 participants had no VR experience, 13 participants had VR experience once or twice, and 7 reported they had experienced VR more than three times. 20 people wore glasses and lenses, and 27 people were right-handed. After the experiment was over, all participants got incentives of 10\$. In order to comply with the government's quarantine rules due to the COVID-19 outbreak, we obtain the user's agreement to transmit experimental-related information to the quarantine authorities, if necessary.
        
\begin{figure}[h]
            \centering
            \subfigure[]{\includegraphics[width=0.49\linewidth]{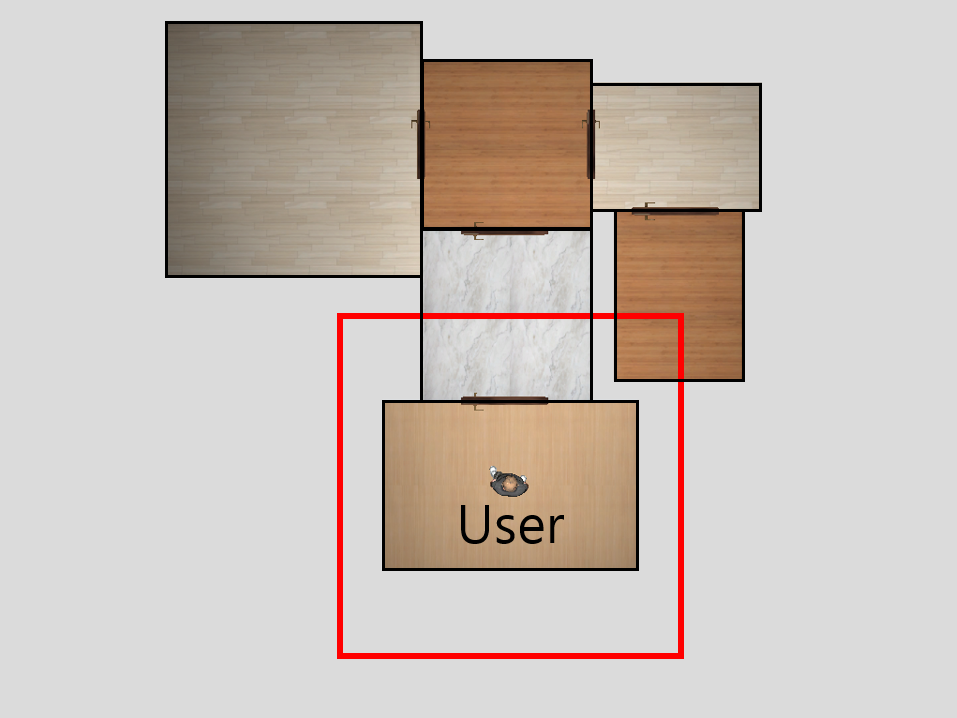}}
            \subfigure[]{\includegraphics[width=0.49\linewidth]{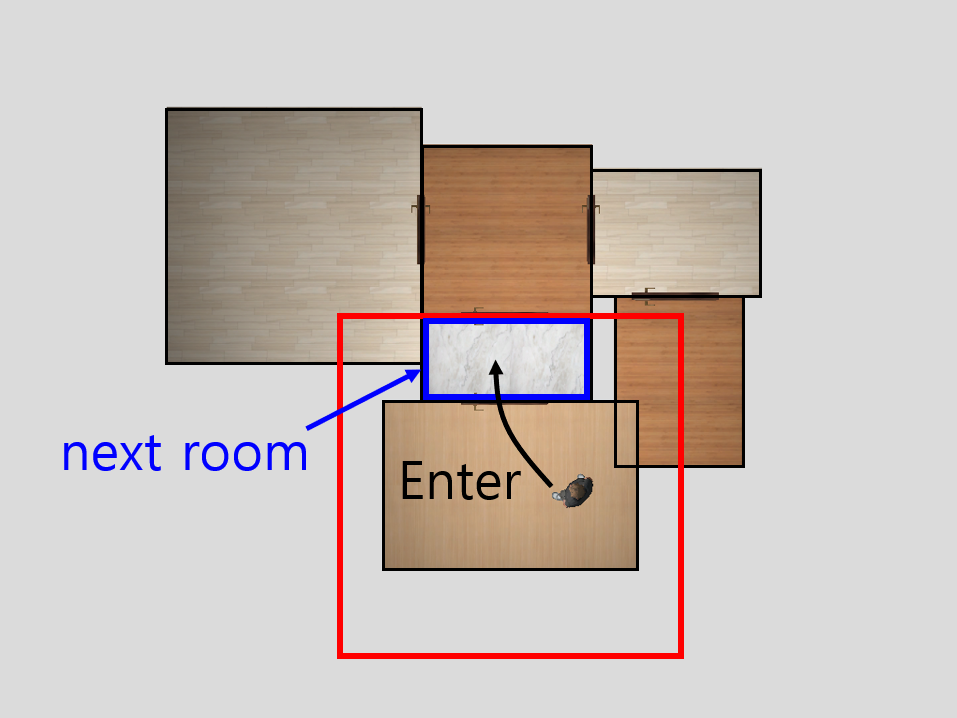}}
            \subfigure[]{\includegraphics[width=0.49\linewidth]{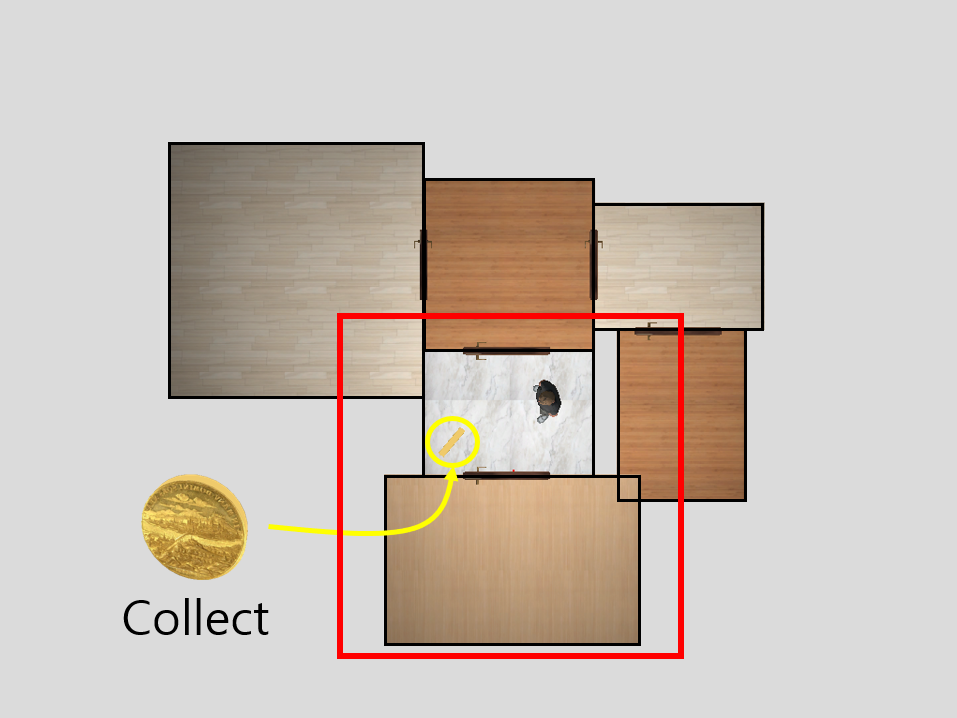}}
            \subfigure[]{\includegraphics[width=0.49\linewidth]{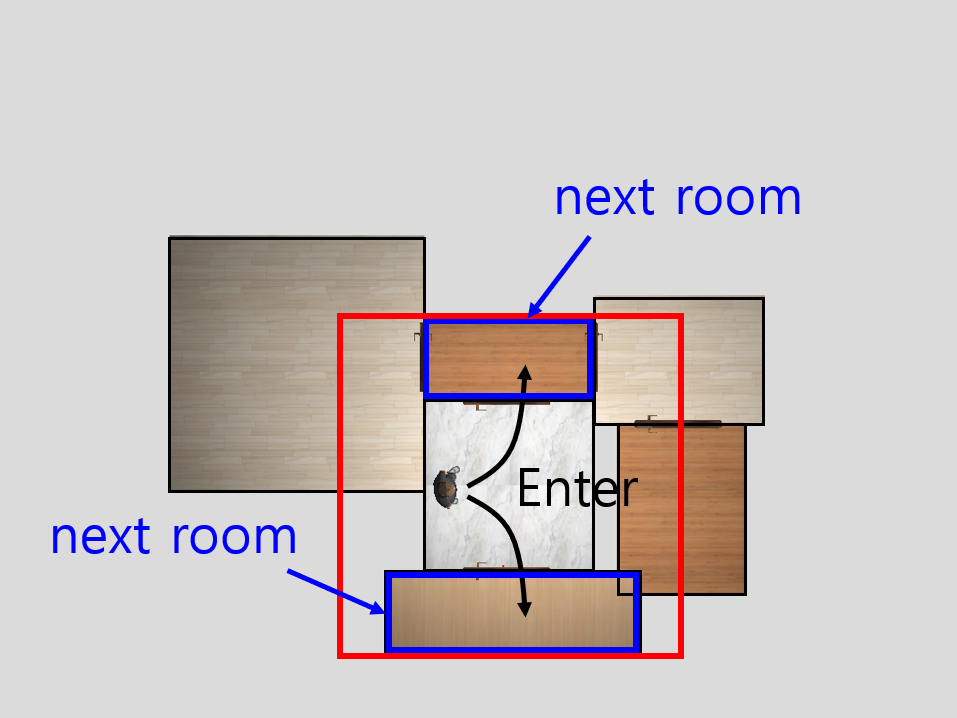}}
            \caption{The entire procedure of Experiment 2: (a) the user initially locates in the center of the room, (b) the user moves to the next room, (c) the user collect coins, (d) when all collecting is done, the user choose next room, move to it, and repeat the above process.}
            \label{fig:exp2}
        \end{figure}

        \paragraph{\bf Procedure}
            We received a preliminary questionnaire from users and collected information such as age and gender for the analysis of the participant group. We conducted a simple test session to prevent the learning effect by adapting the users to locomotion and interaction in VR. After the test session was over, we instructed the user that the tasks to do during the experiment were in progress. As shown in Figure \ref{fig:exp2}(a), the user starts at the center of a virtual room. The user first opened the door and moved to the connected room (Figure \ref{fig:exp2}(b)). Then, as shown in Figure \ref{fig:exp2}(c), the user repeatedly collects coins in the room. If the user collected the given number of coins in the corresponding room, the user chose one of the rooms connected to the current room and repeated the above process. After the given time has elapsed and the experiment is over, the user filled out the SUS, SUSPQ, IEQ, and SSQ questionnaires. Then we changed the experiment condition and the user proceeded with the experiment again. Most of the subjects took about 40 minutes to complete all trials and questionnaires.
        
\begin{figure*}[t]
            \centering
            \subfigure[]{\includegraphics[width=0.4\linewidth]{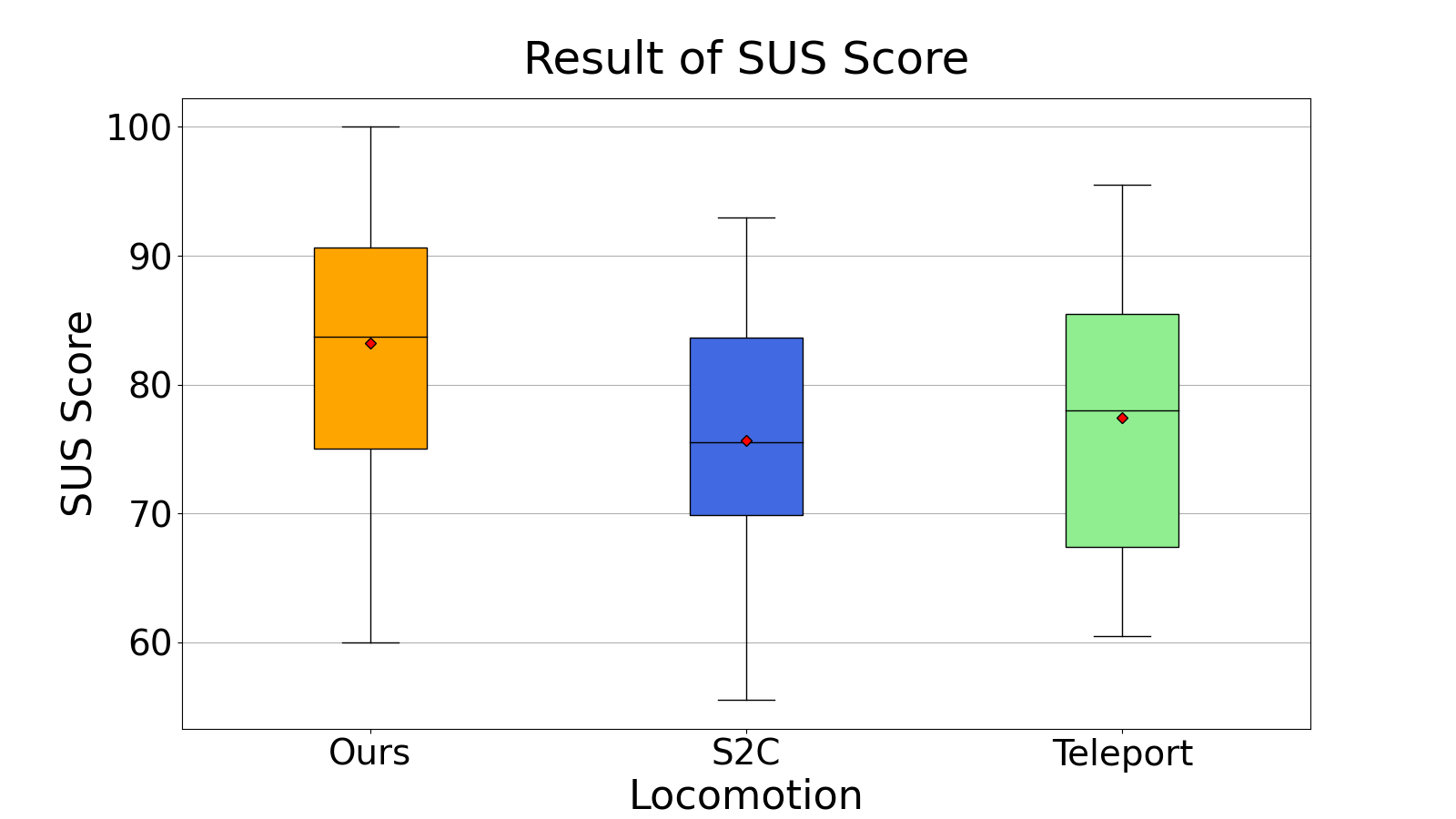}}
            \subfigure[]{\includegraphics[width=0.4\linewidth]{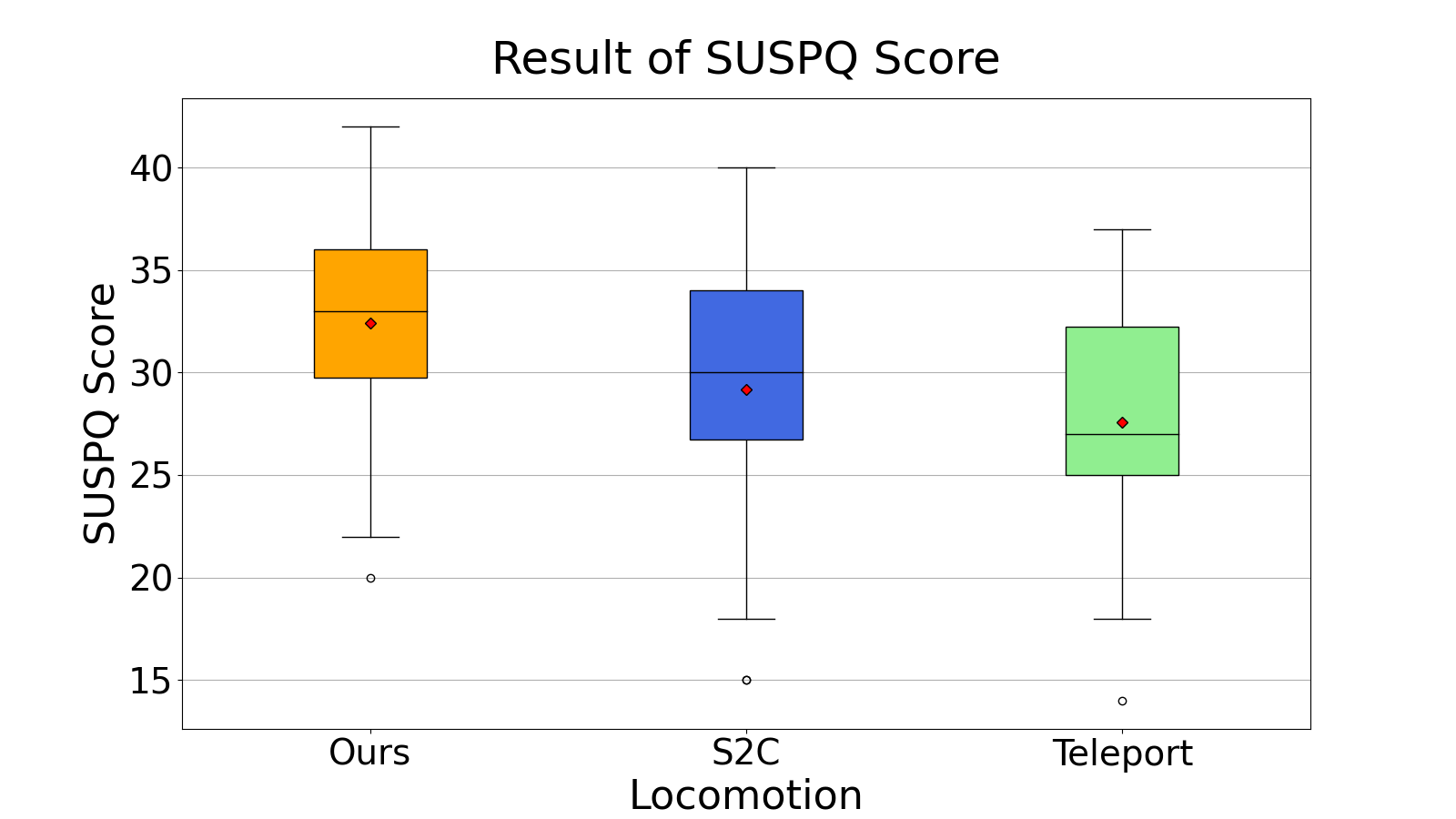}}
            \subfigure[]{\includegraphics[width=0.4\linewidth]{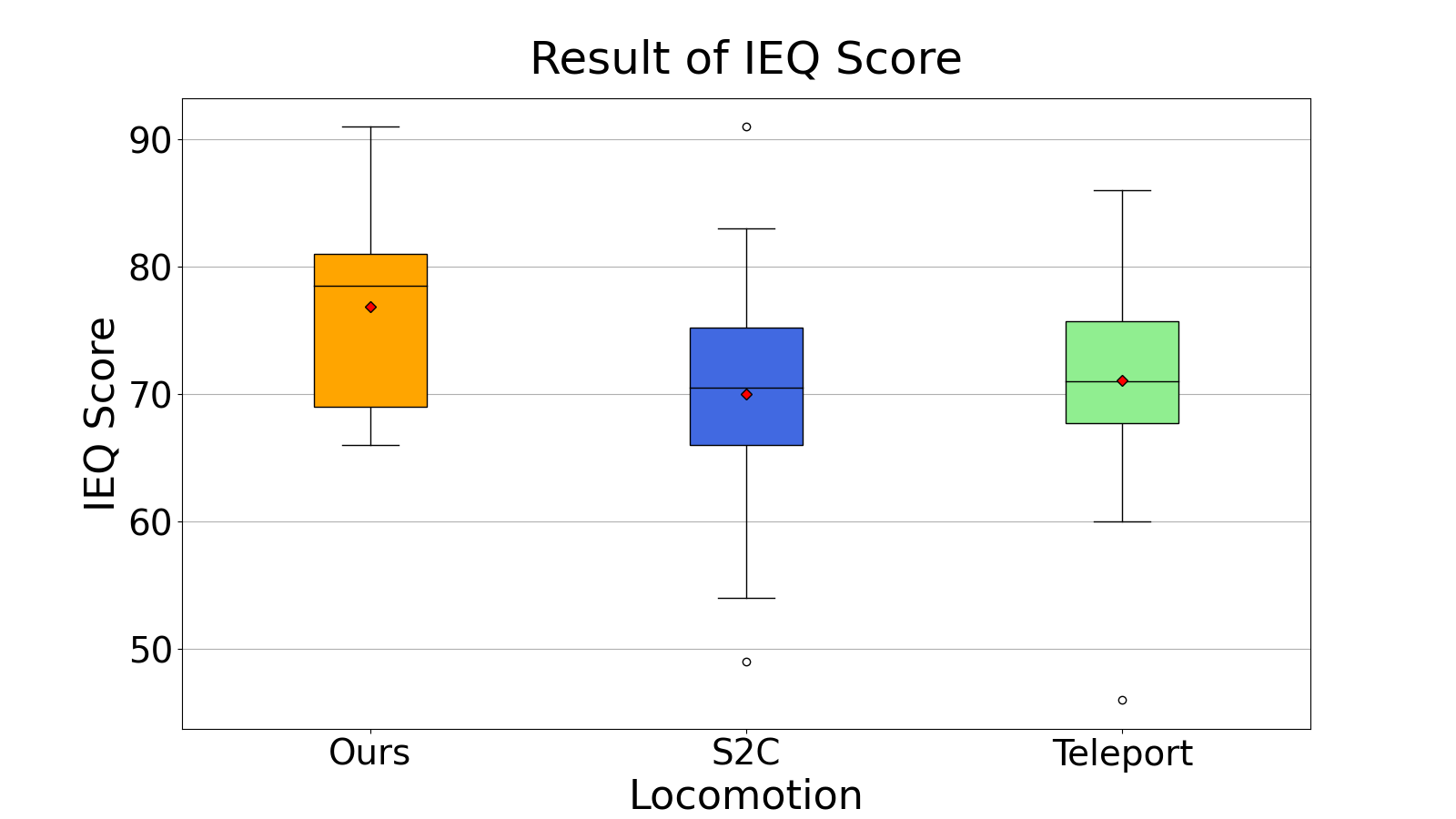}}
            \subfigure[]{\includegraphics[width=0.4\linewidth]{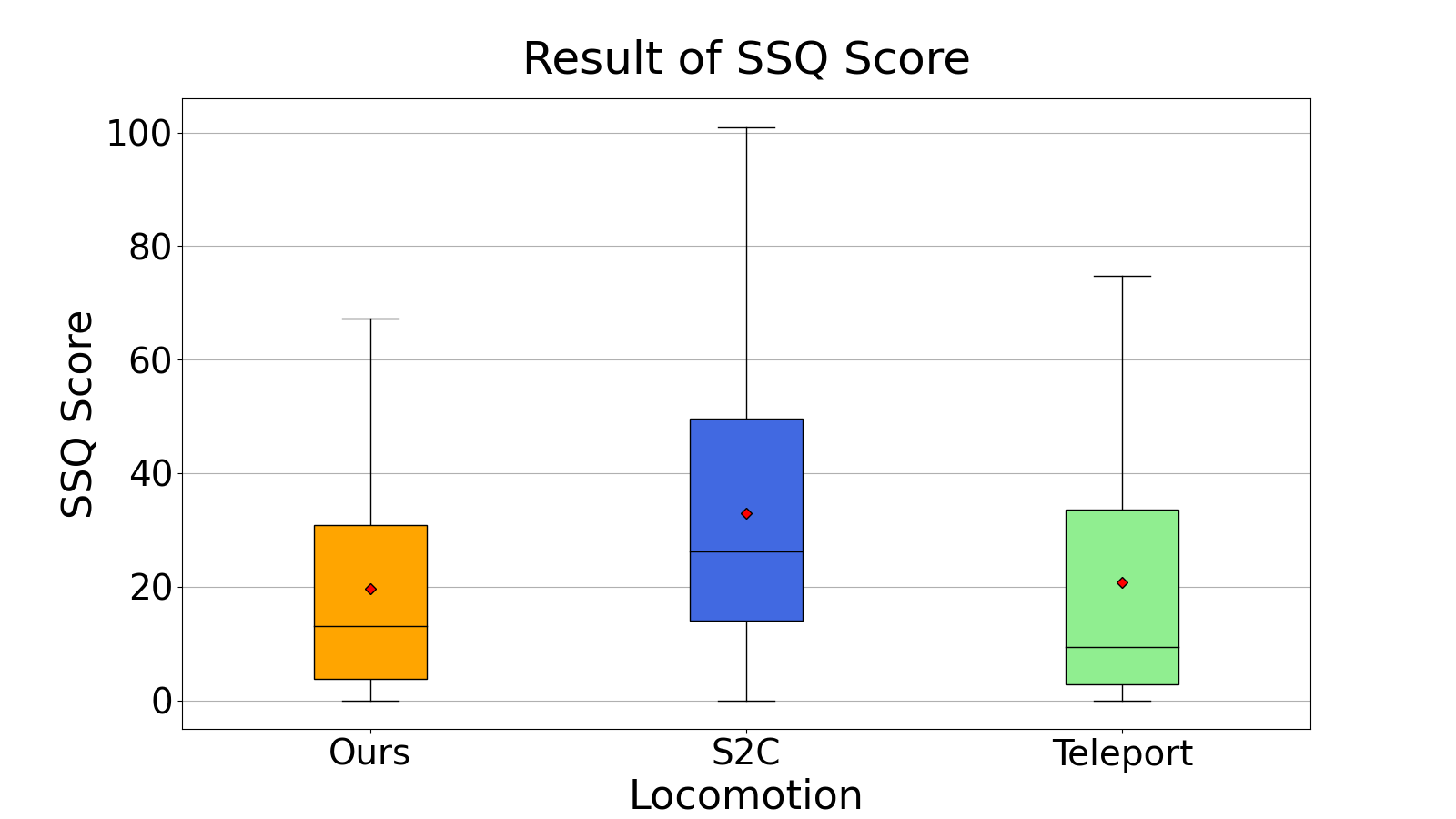}}
            \caption{The box-whisker plot results of user questionnaires for each locomotion: (a) the usability score, (b) the presence score, (c) the immersion score, (d) the motion sickness score.}
            \label{fig:exp2_result}
        \end{figure*}

        \paragraph{\bf Result}
            Table \ref{table:exp2_result} shows the average score of SUS, SUSPQ, IEQ, and SSQ in each locomotion condition. Figure \ref{fig:exp2_result} shows each score as box-and-whisker plot. We used a significance value of $\alpha = .05$ to test all hypotheses. For statistical analysis, we performed a Saphiro-Wilk test [23] for normality and Levene's test for equality of variance on each questionnaire result. All SUS, SUSPQ, and IEQ results passed both normality and equality of variance tests. However, the SSQ result did not pass the normality test. Hence, we performed a one-way analysis of variance (ANOVA) test for SUS, SUSPQ, and IEQ results to check whether each score was significantly different according to locomotion. There were statistically significant differences in SUS ($F(2,84)=4.198835,p=0.01828$), SUSPQ ($F(2,84)=4.351622,p=0.015913$), and IEQ ($F(2,84)=5.752914,p=0.004555$) in each case. A pairwise Tukey’s honestly significant difference test was performed as a posthoc test. As a result, there were significant differences between Ours and S2C ($p=0.0175$) in SUS, Ours and Teleport ($p=0.0131$) in SUSPQ, and Ours and S2C ($p=0.0063$), Ours and Teleport ($p=0.0249$) in IEQ. Meanwhile, since SSQ did not pass the normality test, we conducted the Kruskal–Wallis H test to verify the significant difference in SSQ score among the three methods. we also found a significant difference in SSQ score ($H=11.80002,p=0.0027$) for each method. As a post-hoc test, we conducted the Mann-Whitney U test with adjusting $p$-value ($p_c = 0.0167$ in $5\%$ significance level) by Bonferroni Correction. The result is that there was a significant difference in SSQ scores between Ours and S2C ($p=0.0103$) and S2C and Teleport ($p=0.0092$).
            
        \begin{table}[h]
            \centering
            \begin{tabular}{lm{3em}m{3em}m{3em}m{3em}}
            \toprule
            Method & SUS & SUSPQ & IEQ & SSQ \\
            \midrule
            Ours & \textbf{83.621} & \textbf{32.690} & \textbf{77.000} & \textbf{19.474} \\
            S2C & 75.759 & 29.586 & 70.310 & 36.983 \\
            Teleport & 77.828 & 28.000 & 71.345 & 19.990 \\
            \bottomrule
            \end{tabular}
            \caption{The average scores of user questionnaire measured in all locomotion techniques.}
            \label{table:exp2_result}
        \end{table}

\section{Discussion}
    \subsection{Result of Experiment 1}
        As shown in Table \ref{table:exp1_result} and Figure \ref{fig:exp1_result}, the detection thresholds for wall movement in Long, Middle and Short cases are [0.737,1.211], [0.868,1.200], and [0.899,1.145], respectively. Through this, we confirm that the closer the wall is, the narrower the entire range of detection threshold is.  As mentioned in Experiment 1, it can be said that the change in the user's view increases as the user approaches the wall. Also, we conclude that the amount of change on the user's scene determines to recognize the movement of the wall outside the field of view.  Hence, we can say that the user is more likely to perceive the movement of the nearer wall than that of the further wall. Moreover, as the user gets closer to the wall, we can see that the lower threshold narrows dramatically more than the upper threshold. We believe that this is because the user's point of view is perspective. As a result, the closer the wall is, the greater the change in the user's view is when the wall is closer than away from the user.
        
        On the other hand, we confirmed that the standard deviation of the response rate of expansion for all pairs of gains and distance is more significant than expected, as shown in Figure \ref{fig:exp1_result}(b)(c)(d). This means that although the user's view is affected by the distance between the user and the moving wall due to our experimental results, it also can be affect by many other variables such as the user's viewing angle, the user's current state, the texture of the wall, and the relative location of the surrounding furniture. Therefore, we first express the complexity of the user's view as a set of feature points extracted from the user's scene rather than simply the distance from the wall and then check the detection threshold for wall movement whenever these feature points change.
        
        
    \subsection{Result of Experiment 2}
        In Experiment 2 results, we verified that Ours showed statistically significantly higher performance in all scores except for SUSPQ than S2C. This can be confirmed in Figure \ref{fig:exp1_result}(b)(c)(d). Therefore, we support {\bf H1-1}, {\bf H1-3}, and {\bf H1-4} and reject {\bf H1-2}. This is because the user feels visual-vestibular conflict due to the continuous application of the gain in S2C. It also seems that the resets that occurs during the experiment greatly influence the user's questionnaire. Even in users' feedback in S2C, there were many cases where they felt that the view was rotating during VR, so they felt dizzy. However, we believe that there is no significant difference between Ours and S2C in terms of presence because both methods actually explore the room through walking.
        
        Also, we confirmed that Ours showed statistically significantly higher performance in SUSPQ and IEQ compared to Teleport. Hence, we support {\bf H2-2}, {\bf H2-3} and reject {\bf H2-1}, {\bf H2-4}. 
        In the case of usability, it seems that once the user gets used to the Teleport method, they can move conveniently using Teleport in the same place. In short, users could not adapt to the Teleport method at first, but they got used to it during the test session and can move conveniently in place. Hence, it seems that it did not show a big difference in usability compared to Ours. Looking at the user feedback about Teleport, some users believed that it was a little uncomfortable because it took time to adapt to Teleport. Meanwhile, there was feedback that the feeling of immersion was rather increased because Teleport was an impossible experience in reality. In the case of motion sickness, it seems that the user feels motion sickness unconsciously in Ours method, because the room's structure continues to change while exploring the room due to the constant room manipulation of Ours.
        
    \subsection{Limitation}
        Our proposed method has some limitations: the user has to stay in the room and look around continuously until the algorithm completes restore-compression phases. Briefly, the user cannot move across multiple rooms in a row because the user should perform a certain task every time he/she visits each room. Also, since this method assumes that the virtual room and the real space have a rectangular shape, the algorithm is currently not suitable in a virtual indoor environment having general polygonal or circular rooms. Similarly, our method does not allow the environment to have a rectangular room that does not fit entirely inside the real space. 
        Also, if the size of a compressed room is too small, the user cannot practically explore that room on foot and compromise the sense of immersion. As a result, the extent to which it can be compressed is somewhat limited. In other words, the size of each room is bound to be much smaller than the given tracking space to ensure a minimum size when neighboring rooms are compressed. Besides, to use change blindness, we are currently only using wall movement that is wholly outside the user's field of view. For this reason, only specific walls can move. 
        Therefore, there is a need to find a way to alleviate this limitation of freedom by moving the walls inside the field of view as well as the walls outside the user's field of view.
        
        We newly defined the wall movement gain that can be applied in a general virtual room surrounded by walls and measured the detection threshold of the wall and distance. However, it is necessary to measure the detection threshold for out-of-sight wall movement gain more accurately by considering not only the distance from the wall but also other variables. In particular, since change blindness is directly related to the change in the user's view, we need to quantitatively analyze the user's viewing point and estimate the detection threshold for the movement of the wall according to this value.

\section{Conclusion}
    In this study, 
    we propose a new spatial manipulation technique in RDW by moving the wall outside the field of view using change blindness applicable to various virtual indoor environments that can be represented as infinitely rooms are interconnected. The method consists of restore and compression phases. The restore phase continuously moves the walls outside the user's field of view using change blindness. After the restoration, the room where the current user is located has the original dimension of the room and is located in the center of the real space. The compression phase pushes all rooms adjacent to the restored room into the real space. In addition, we defined the outside wall movement to utilize change blindness as a new type of gain and measured its detection threshold. We can manipulate the space by applying this gain to the typical virtual room surrounded by walls. In addition, we conducted a live user experiment between the proposed method, S2C, and Teleport and surveyed users after the experiment.
    
    Our future studies that can be conducted from this work are as follows. First, we extend this technique to a more general indoor environment with a virtual unit that is not rectangular but has various shapes and a larger size than the real space. Also, our current method only uses out-of-sight wall movement. To increase the degree of freedom in spatial manipulation, we can consider the extension of the algorithm to move some objects and walls inside the user's view in some way. 
    On the other hand, by applying reinforcement learning, it will be possible to determine the room's optimal restored size and location in the restoration phase according to the state of the user and environment. Finally, we can consider combining our method with the previously studied techniques such as flexible space \cite{vasylevska2013flexible} and impossible space \cite{suma2012impossible}. For example, there can be several themes in a virtual museum such as ``Stone Age" and ``Bronze Age". We can use our method to navigate several rooms in an area of a certain single theme. Then, using the flexible space technique \cite{vasylevska2013flexible} we can construct the way to connect a theme area to another theme area by procedurally creating the corridor and rooms. Then users do not have to do unnecessary tasks when simply moving between the areas of different themes.  


\bibliographystyle{abbrv-doi}

\bibliography{05revision_template.bbl}
\end{document}